\documentclass[aps,prb,superscriptaddress,twocolumn,showpacs,a4paper,footinbib]{revtex4-1}
\usepackage[breaklinks=true,bookmarks=false,colorlinks=true,citecolor=violet,urlcolor=blue!60!black,linkcolor=red]{hyperref}
\usepackage{amsfonts,amsmath}
\usepackage{graphicx}
\usepackage{color}
\usepackage{tikz}
\usetikzlibrary{shapes,decorations}
\usepackage{braket}

\usetikzlibrary{arrows,decorations.pathmorphing,backgrounds,positioning,fit,petri,shapes,calc,decorations.markings}
\begin{document}
\title{Shannon and R\'enyi mutual information in quantum critical spin chains}
\author{Jean-Marie St\'ephan}
\affiliation{Physics Department, University of Virginia, Charlottesville, VA 22904-4714}
 \date{\today}
\begin{abstract}
 We study the Shannon mutual information in one-dimensional critical spin chains, following a recent conjecture (Phys. Rev. Lett. {\bf 111}, 017201 (2013)[\onlinecite{AlcarazRajabpour}]), as well as R\'enyi generalizations of it. We combine conformal field theory arguments with numerical computations in lattice discretizations with central charge $c=1$ and $c=1/2$. For a periodic system of length $L$ cut into two parts of length $\ell$ and $L-\ell$, all our results agree with the general shape-dependence $I_n(\ell,L)=(b_n/4)\ln \left(\frac{L}{\pi}\sin \frac{\pi \ell}{L}\right)$, where $b_n$ is a universal coefficient. For the free boson CFT we show from general arguments that $b_n=c=1$. At $c=1/2$ we conjecture a result for $n>1$. We perform extensive numerical computations in Ising chains to confirm this, and also find $b_1\simeq 0.4801629(2)$, a nontrivial number which we do not understand analytically. Open chains at $c=1/2$ and $n=1$ are even more intriguing, with a shape-dependent logarithmic divergence of the Shannon mutual information. 
\end{abstract}
 \maketitle

\section{Introduction}
The entanglement entropy (EE) has emerged as a particularly convenient tool in the study of quantum many-body systems. For example in one-dimensional critical systems it is known to be universal \cite{EE1d1,EE1d2,EE1d3}, with a slow logarithmic divergence proportional to the central charge of the underlying conformal field theory \cite{BPZ,Yellowpages}(CFT). It is not the only information-theoretic quantity that exhibits universal behavior: various types of other entanglement\cite{negativity} or fidelity\cite{LBF} measures have been shown to be universal. However one particularly attractive feature of the EE is that it only depends to leading order on the central charge, not on more refined properties of the CFT. 

The Shannon entropy, a measure of disorder in a certain basis, is also an interesting quantity in its own right. It is defined as
\begin{equation}
 S=-\sum_{\sigma} p_\sigma \ln p_\sigma\qquad,\qquad p_\sigma=\left|\braket{\sigma|\psi}\right|^2,
\end{equation}
where $\ket{\psi}$ will be for us the ground state wavefunction. The sum runs over all configurations in the chosen basis of the Hilbert space. This entropy has been studied in a variety of contexts, from the Anderson localization \cite{ErversMirlin,ErversMirlinPRL}, multifractality\cite{AtasBogomolny} and quantum chaos \cite{Sinai,Izrailev}, to quantum quenches\cite{SantosRigol} and critical phenomena. 
It is is also related \cite{SFMP} to the entanglement entropy of certain 2d Rokhsar-Kivelson states \cite{FradkinMoore,Hsuetal,SFMP,SMP2,Oshikawa,Zaletel,SMP3,RVB1,RVB2}, as well as the classical mutual information in 2d systems \cite{GMI}. 

In the following we will only consider a basis obtained from tensor products of local degrees of freedom.    
Typically in such cases the Shannon entropy obeys a ``volume'' law\footnote{The volume law may be violated if the locality requirement is dropped. For example the Shannon entropy evaluated in the basis of the eigenstates of the Hamiltonian --which is usually highly nonlocal-- is $0$ if the ground-state is nondegenerate.}: it is proportional to the volume $L^d$ of the $d-$dimensional quantum system. Subleading terms are also interesting, as they have been shown to be universal. For example in periodic 1d chains whose low-energy properties are governed by a Luttinger liquid theory, a subleading term gives access to the Luttinger parameter \cite{SFMP}. In other systems or geometries it can be used to extract universal properties \cite{SMP2,Zaletel,SMP3} of the underlying CFT, and thus identify it. We refer to Ref.~[\onlinecite{AletShannon3}] for a review. Universal terms can also be accessed in higher dimensions \cite{AletShannon,AletShannon2,AletShannon3}.

Inspired by all the results obtained for the one-dimensional entanglement entropy, it is natural to study the Shannon entropy of a subsystem \cite{Umetal}. Let us cut our chain in two parts $A$ and $B$. Subsystem $A$ is described by the reduced density matrix $\rho_A={\rm Tr}_B \ket{\psi}\bra{\psi}$, and the subsystem Shannon entropy is
\begin{equation}
 S(A)=-\sum_{\mu} p_\mu \ln p_\mu\qquad,\qquad p_\mu=\braket{\mu|\rho_A|\mu}.
\end{equation}
Alternatively, $p_\mu$ can be seen as the marginal probability obtained from summing over all spin configurations outside of $A$. Then, a most natural object is the mutual information between them, defined as
\begin{equation}\label{eq:smidef}
 I(A,B)=S(A)+S(B)-S(A\cup B).
\end{equation}
$S(A\cup B)$ is the Shannon entropy of the total system. 
Note that the mutual information is symmetric with respect to $A$ and $B$, and that the leading contributions proportional to $L$ cancel in the definition (\ref{eq:smidef}). 
It has been first studied for the Ising universality class \cite{Umetal,LauGrassberger}, and the scaling argued to be universal. More precisely, it was found that the Shannon  mutual information (SMI) is well described by the following formula:
\begin{equation}\label{eq:smiscaling}
 I(\ell,L)=\frac{b}{4}\ln \left(\frac{L}{\pi}\sin \frac{\pi \ell}{L}\right)+O(1),
\end{equation}
for a periodic system of length $L$ cut into two parts of respective lengths $\ell$ and $L-\ell$.
The $O(1)$ term includes a non-universal constant as well as subleading corrections. 
This scaling form is almost identical to the celebrated EE result \cite{EE1d1,EE1d3}, upon substituting $c/3$ with $b/4$. In the following, we will refer to such a scaling as the \emph{conformal scaling}.
In an interesting recent development, Alcaraz and Rajabpour \cite{AlcarazRajabpour} further conjectured that the coefficient $b$ appearing in (\ref{eq:smiscaling}) is nothing but the central charge $c$ of the underlying CFT in general. This conjecture was supported by exact diagonalizations in a variety of spin chains corresponding to different universality classes, as well as on an exact correspondence with the second R\'enyi entanglement entropy for a particular model of harmonic oscillators with $c=1$.  

This conjecture, while very reasonable, is intriguing for two reasons. First, there is a small mismatch (of the order of $2$ percent) between the numerical estimate of $b$ in Ref.~[\onlinecite{AlcarazRajabpour}] compared to  Refs.~[\onlinecite{PhDStephan,Umetal,LauGrassberger}], for different models and limits belonging to the Ising universality class. On the conceptual level also, the result appears much simpler than previous findings in the Shannon entropy of the full chain \cite{SFMP,SMP2,Zaletel}. Indeed, highly nontrivial transitions in the R\'enyi entropy,
\begin{equation}
 S_n=\frac{1}{1-n}\ln \left(\sum_\sigma \left[p_{\sigma}\right]^n\right),
\end{equation}
were observed as a function of the R\'enyi index $n$. For Ising, this transition occurs precisely at $n=1$, where $S_n$ reduces to the Shannon entropy $\lim_{n\to 1}S_n=S_1=S$. 

The above considerations motivate us to clarify this issue and revisit this problem, generalizing the study to that of the R\'enyi mutual information (RMI)
\begin{equation}\label{eq:rmidef}
 I_n(A,B)=S_n(A)+S_n(B)-S_n(A\cup B)
\end{equation}
for general $n$. A unifying conclusion from our study will be that the RMI of periodic systems generically obeys the conformal scaling
\begin{equation}\label{eq:general_rmi}
 I_n(\ell,L)=\frac{b_n}{4}\ln\left(\frac{L}{\pi}\sin \frac{\pi \ell}{L}\right)+O(1),
\end{equation}
where $b_n$ is a \emph{universal} -- and in general nontrivial -- coefficient. We also explain why such a result should be, typically, independent of the choice of local basis. We derive Eq.~(\ref{eq:general_rmi}) for the free boson CFT, exploiting the gaussian form of the action, and combining this with boundary CFT arguments. We show that $b_n$ is proportional to the central charge in this case. At $n=1$ we recover the numerical result of Ref.~[\onlinecite{AlcarazRajabpour}]. Crucial to our derivation is the gaussian nature of the action. 

Already for the Ising CFT such arguments do not apply anymore. To investigate the scaling of the RMI we performed extensive numerical simulations on the example of the XY chain in transverse field. We used the free fermion structure to access larger system sizes than considered in previous works, and confirm (\ref{eq:general_rmi}). We then argue for a formula proportional to the central charge when $n>1$. Using the very accurate free fermion data and combining with simple extrapolation techniques, we also managed to obtain
\begin{equation}
 b=b_1=0.4801629(2)
\end{equation}
at the Shannon point. This differs from the central charge ($c=1/2$), and convincingly disproves the general conjecture of Ref.~[\onlinecite{AlcarazRajabpour}]. However the conformal scaling still holds. We do not know how to derive this from first principles, but conjecture that all minimal models give similar results to that of Ising, with (Shannon) prefactor close but not quite identical to the central charge. 

The paper is organized as follows. In Sec.~\ref{sec:CFT} we study the path-integral representation of the R\'enyi mutual information. We derive the conformal scaling for the free boson CFT ($c=1$), and show that (\ref{eq:general_rmi}) holds (with $b_n=1$) for all R\'enyi entropies, provided $n$ is not too large. We also show how the phase transition scenario of Ref.~[\onlinecite{SMP2,Zaletel}] allows us to derive the scaling of the mutual information for the Ising universality class provided $n>1$. These predictions are checked numerically in Sec.~\ref{sec:numerics} for two spin chains, the XXZ and XY chain in transverse field, where very good agreement is found. Sec.~\ref{sec:marginal} focuses on a numerical extraction of the universal coefficient $b_n$ in free fermionic systems, as well as on the ``transition'' points where our analytical arguments do not apply. Some additional details are gathered in three appendices. The first shows how the conformal scaling can be derived from standard boundary CFT techniques (App.~\ref{sec:mappings}). The second (App.~\ref{sec:correlations}) shows some numerical computations of correlation functions, in support of our arguments for Ising. Finally, we perform in App.~\ref{sec:exact} some exact computations of R\'enyi entropies for integer $n$ in the full XX chain, one of the simplest models described by a free boson CFT.
\section{Mutual information in CFT}
\label{sec:CFT}
In this section we study the R\'enyi mutual information for all R\'enyi indices $n$ in a CFT setup. 
\subsection{Infinite Renyi limit}
\label{sec:efp}
\begin{figure}[htbp]
\includegraphics[width=3.2cm]{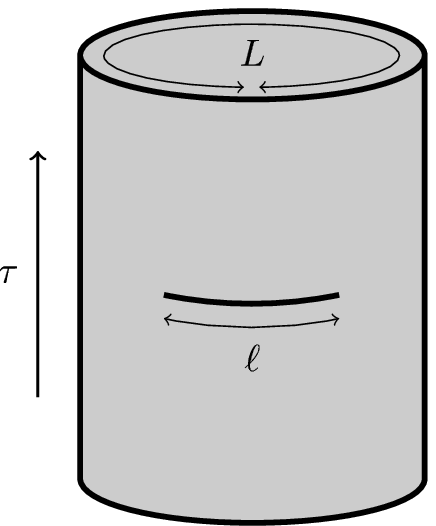}\hfill
\includegraphics[width=3.2cm]{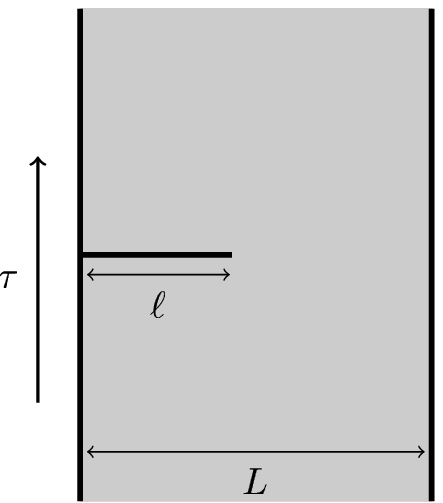}
\caption{Space-time geometry. (a) cut-cylinder, corresponding to a periodic chain. (b) cut-strip, corresponding to an open chain. By convention the position of the slit is at $\tau=0$ and $x\in [0;\ell].$}
 \label{fig:geometry}
\end{figure}

Let us start by considering the case $n\to \infty$. This limit turns out to be simplest, and will be useful later on. The entropy is given by
\begin{equation}
 S_\infty(A)=-\ln \left(\braket{\textrm{max}|\rho_A|\textrm{max}}\right).
\end{equation}
A similar expression holds for $S_\infty(B)$ and $S_{\infty}(A\cup B)$. Here $\ket{\textrm{max}}$ denotes the spin configuration(s) with the biggest probability(ies). In most spin chains these are usually attained by homogeneous states. For example in the (ferromagnetic) Ising chain in transverse field $\ket{\textrm{max}}=\ket{\uparrow\ldots\uparrow}$ or $\ket{\textrm{max}}=\ket{\downarrow\ldots\downarrow}$, and in the antiferromagnetic XXZ chain $\ket{{\rm max}}=\ket{\uparrow\downarrow\ldots\uparrow\downarrow}$ or $\ket{{\rm max}}=\ket{\downarrow\uparrow\ldots\downarrow\uparrow}$. In a euclidean picture this probability is given by
\begin{equation}
 \braket{\textrm{max}|\rho_A|\textrm{max}}=\lim_{\tau\to \infty}\frac{\braket{a|e^{-\tau H} \delta (\sigma-\sigma_{\rm max})e^{-\tau H}|a}}{\braket{a|e^{-2\tau H}|a}}
\end{equation}
$H$ is the Hamiltonian of the system, $\ket{a}$ is a state at infinity that has non-zero overlap with the ground state, and the $\tau \to \infty$ limit ensures the projection onto the ground-state of the Hamiltonian $H$. $\delta(\sigma-\sigma_{\rm max})$ selects the spin configuration(s) with maximum probability(ies) on the segment $\tau=0,0\leq x\leq \ell$. In a transfer matrix picture, this can be represented as the following ratio of partition functions
\begin{equation}
  \braket{\textrm{max}|\rho_A|\textrm{max}}=\frac{\mathcal{Z}_{\rm slit}}{\mathcal{Z}}.
\end{equation}
For a periodic (open) system ${\cal Z}_{\rm slit}$ is the partition function of an infinite cylinder (strip) with a slit, as is shown in Fig.~\ref{fig:geometry}(a,b). 
The homogeneous sequence of spins imposed on the slit in Fig.~\ref{fig:geometry} should renormalize to a conformal invariant boundary condition (CIBC) \cite{Cardybcc}, and such an observation allows for a standard boundary CFT treatment. In fact, this exact ratio has already been studied in Ref.~[\onlinecite{emptiness}], under the name emptiness formation probability. Denote by $\mathcal{E}=-\ln \left(\mathcal{Z}_{\rm slit}/\mathcal{Z}\right)=S_\infty(A)$ the logarithm of this ratio. For a periodic system, we have, to the leading order \cite{emptiness}
\begin{equation}\label{eq:efp_periodic}
 \mathcal{E}_p=a_1\ell+\frac{c}{8}\ln \left[\frac{L}{\pi}\sin\left(\frac{\pi \ell}{L}\right)\right]+O(1).
\end{equation}
$a_1$ is a non-universal line free energy and $c$ is the (universal) central charge of the CFT. For the sake of completeness, we recall the derivation of this result in Appendix \ref{sec:mappings}. In (\ref{eq:efp_periodic}), the coefficient of the logarithmic divergence $\ln L$ is a direct consequence of the presence of sharp corners in the geometry \cite{CardyPeschel}. Indeed, for each corner with angle $\theta$ there is a contribution
\begin{equation}
 \Delta F=\left[\frac{c}{24}\left(\frac{\theta}{\pi}-\frac{\pi}{\theta}\right)+h\frac{\pi}{\theta}\right]\ln L
\end{equation}
to the free energy. $h$ is the dimension of a boundary changing operator \cite{Cardybcc}, in case there are changes in CIBC. This is possible in the strip geometry (see Fig.~\ref{fig:geometry}(b)), but not on the cylinder. Hence $h=0$ and the coefficient of the contribution, $c/8\ln L$, follows from the $2$ corners with angle $2\pi$ in Fig.~\ref{fig:geometry}(a). The total entropy $S_\infty(A\cup B)$ contributes to an universal $O(1)$ term, which does not play a role. 
Using (\ref{eq:efp_periodic}) for both $S_\infty(A)$ and $S_\infty(B)$, we get for the mutual information
\begin{equation}
 I_{\infty}(\ell,L)=\frac{c}{4}\ln \left[\frac{L}{\pi}\sin \left(\frac{\pi \ell}{L}\right)\right]+O(1).
\end{equation}
The open geometry is similar. The homogeneous configuration of spins will also renormalize to a CIBC, but this may differ from the CIBC at the external boundary. We obtain in the most general case
\begin{equation}\label{eq:efp_open}
 \mathcal{E}_o=a_1\ell +\frac{c}{16}\ln \left[\frac{L}{\pi}\sin \frac{\pi \ell}{L}\right]+\left(4h-\frac{c}{8}\right)\ln\left[\frac{L}{\pi}\tan\frac{\pi \ell}{2L}\right],
\end{equation}
where $h$ is the dimension of the possible boundary changing operator. There are also four $\pi/2$ corners (see Fig.~\ref{fig:geometry}(b) when $\ell=L$) in the R\'enyi entropy of the full chain $S_\infty(A\cup B)$, that contribute to a $(-c/4+8h)\ln L$. Adding up all contributions, the mutual information becomes
\begin{equation}\label{eq:Iinfty}
 I_{\infty} (\ell,L)=\frac{c}{8}\ln \left[\frac{L}{\pi}\sin \left(\frac{\pi \ell}{L}\right)\right]+O(1).
\end{equation}
This is exactly half the periodic result. Interestingly, it does not depend on the conformal dimension $h$, as the corresponding part is antisymmetric in (\ref{eq:efp_open}). Therefore the precise nature of the boundary conditions plays no role in the RMI, \emph{provided} it is conformal. This will be very important in the following, as a change of basis can typically change also the CIBC. It is therefore reasonable to hope that the RMI will be robust to such changes.  

Let us also mention that subleading corrections can be computed in these two simple cases. The leading corrections \cite{logloverl,emptiness} to (\ref{eq:efp_periodic},\ref{eq:efp_open}) take the form of a series in $\beta_k(\ln L)L^{-k}$, where $\beta_k(x)$ is a polynomial of degree at most $k$ in $x$.
\subsection{The free boson case}
\label{sec:freebosoncft}
This section is devoted to the free compact field with action
\begin{equation}\label{eq:action}
 {\cal A}=\frac{g}{4\pi}\int_{0}^{L} dx \int_{-\infty}^{\infty} d\tau \left(\nabla \varphi\right)^2\quad,\quad \varphi\equiv \varphi+2\pi r. 
\end{equation}
The field $\varphi$ is compactified on a circle of radius $r$. This is the euclidean time version of the Luttinger liquid CFT; the Luttinger parameter, that controls the decay of correlation functions, is $K=(2gr^2)^{-1}$. For a periodic (open) chain the field lives on a  infinite cylinder (strip) of circumference (width) $L$.  In the continuum limit, each spin configuration $\mu$ in subsystem $A$ is replaced by a corresponding field configuration
\begin{equation}
\phi(x)=\varphi(x,\tau=0)\qquad,\qquad 0 \leq x\leq \ell.
\end{equation}
The Renyi entropy is
\begin{equation}
 \mathcal{S}_n=\frac{\ln {\cal Z}^{(n)}}{1-n}\qquad,\qquad \mathcal{Z}^{(n)}=\int \left[{\cal D}\phi\right] \left(p_g(\phi)\right)^n.
\end{equation}
We kept track of the stiffness $g$, for reasons that will become apparent shortly.  
The probability $p_g(\phi)$ can be evaluated following Ref.~[\onlinecite{SMP3}]. We decompose the field $\varphi$ into the sum of two terms. The first is a harmonic function $\varphi_\phi$ that satisfies the boundary condition
\begin{equation}
\varphi_\phi(x,\tau=0)=\phi(x)\qquad ,\forall x\in [0;\ell],                                                                                                                                                                                                                                                                                                                                                                                               \end{equation}
and the second an oscillator part $\varphi_0$ that satisfies a Dirichlet boundary condition. We now exploit the gaussian nature of the action, as well as the fact that $\varphi_\phi$ has a vanishing laplacian, to get $\mathcal{A}[\varphi_\phi+\varphi_0]=\mathcal{A}[\varphi_\phi]+\mathcal{A}[\varphi_0]$. Hence
\begin{equation}\label{eq:proba}
 p_g(\phi)=\exp(-S_g[\varphi_\phi])\frac{\mathcal{Z}_g^D}{\mathcal{Z}_g}.
\end{equation}
$\mathcal{Z}_g$ is the partition function of an infinite cylinder (resp. strip), and $\mathcal{Z}^D$ is the partition function of an infinite cylinder (resp. strip) with a Dirichlet defect line at $\tau=0$, $0\leq x\leq \ell$, just as in Fig.~\ref{fig:geometry}.
Now we raise $p_g(\phi)$ to the power $n$, to get 
\begin{equation}\label{eq:proban}
 \left[p_g(\phi)\right]^n=\exp(-n S_g[\varphi^\phi])\left(\frac{\mathcal{Z}_g^D}{\mathcal{Z}_g}\right)^n.
\end{equation}  
The crucial point is that the exponential prefactor in (\ref{eq:proban}) can be interpreted as a Boltzmann factor in a system with stiffness $g^\prime=ng$. Hence we get
\begin{equation}\label{eq:changedstiffness}
 [p_g(\varphi^\phi)]^n\propto p_{ng}(\varphi^\phi).
\end{equation}
The R\'enyi index $n$ therefore \emph{changes} the stiffness to $g^\prime=ng$ (or alternatively, the Luttinger parameter to $K^\prime=K/n$) near the slit. Keeping track of the proportionality coefficients and using the normalization of the $p_{ng}$, we finally arrive at 
\begin{equation}\label{eq:smp}
 {\cal Z}^{(n)}=\frac{\mathcal{Z}_{ng}}{\mathcal{Z}_{ng}^D}\left(\frac{\mathcal{Z}_g^D}{\mathcal{Z}_g}\right)^n.
\end{equation}
This result generalizes Ref.~[\onlinecite{SMP3}] to an arbitrary subsytem of length $\ell$. 

A crucial property of the two cut-cylinder (a) and cut-strip (b) geometries we focus on is that they can be conformally mapped to the upper half-plane (see. \ref{sec:mappings}). 
The leading universal shape-dependent piece is exactly that given in \ref{sec:efp}, and the result only depends on the central charge $c$, \emph{not} the stiffness $g$ \footnote{Note that this is not true for the full periodic chain, where there are no logarithms, and the leading universal piece depends on the stiffness \cite{SFMP,SMP3}.}. Hence the stiffness dependence in the partition functions can be discarded, and we get
\begin{equation}\label{eq:freebosonZreplicas}
 \mathcal{Z}^{(n)}=\left(\frac{\mathcal{Z}^D}{\mathcal{Z}}\right)^{n-1},
\end{equation}
so that
\begin{equation}
 \mathcal{S}_n=-\ln \left(\mathcal{Z}^D/\mathcal{Z}\right).
\end{equation}
This ratio is formally identical to the one studied in the previous section. Therefore we obtain
\begin{equation}\label{eq:freeboson}
 I_n(\ell,L)=\frac{c}{4}\ln \left[\frac{L}{\pi}\sin \frac{\pi \ell}{L}\right]+O(1)\quad,\quad c=1
\end{equation}
for a periodic system, and half that in a open system. The result (\ref{eq:freeboson}) should also apply to orbifolds of the free boson theory, as well as in the non compact limit $r\to\infty$.
When $n=1$ our derivation recovers the numerical results of Ref.~[\onlinecite{AlcarazRajabpour}] for the $XXZ$ chain and the $Q=4$ state quantum Potts model. We emphasize that even though the central charge appears in Eq.~(\ref{eq:freeboson}), the derivation relies crucially on Eq.~(\ref{eq:smp}), which is specific to the free field ($c=1$).

The above derivation implicitly assumes that the boundary is still critical at stiffness $g^\prime=ng$, and this may not necessarily be so. For example lattice effects may change the result if $n$ gets too large in a compact theory. These effects can be tackled by adding vertex operators to the action 
(\ref{eq:action}). The least irrelevant is $V_d=\cos( \frac{d}{r}\varphi)$, where $d$ is the smallest integer allowed by the lattice symmetries. It is irrelevant provided $d^2>2gr^2=K^{-1}$, which is the case since we are studying a critical system. However, in presence of a stiffness $ng$ this condition becomes $d^2>2ngr^2=nK^{-1}$, which can be rewritten as \cite{SMP3}
\begin{equation}\label{eq:nc}
 n<n_c\qquad,\qquad n_c=K d^{\,2}.
\end{equation}
Therefore, when $n>n_c$ a phase transition takes place at the boundary, and the field gets locked into one of the $d$ minima of the cosine potential $V_{d}$. In this boundary-locked phase the universal contributions to the entropy are given by
\begin{equation}
 S_n\sim \frac{1}{1-n}\ln \left[d (p_{max})^n\right],
\end{equation}
where $p_{\rm max}=\braket{\rm max|\rho_A|\rm max}$. The universal shape-dependence becomes
\begin{equation}
 \mathcal{S}_{n>n_c}=\frac{n}{n-1}\ln \left(\frac{\mathcal{Z}^D}{\mathcal{Z}}\right).
\end{equation}
Using the result established in Sec.~\ref{sec:efp}, we get
\begin{equation}\label{eq:freeboson_afternc}
 I_{n>n_c}(\ell,L)=\frac{n}{n-1}\frac{c}{4}\ln \left[\frac{L}{\pi}\sin \frac{\pi \ell}{L}\right]+O(1)
\end{equation}
for a periodic system, and once again half that in a open system. We recover the result Eq.~(\ref{eq:Iinfty}) in the limit $n\to \infty$. Interestingly the conformal scaling holds in both phases ($n<n_c$) and ($n>n_c$), but with different prefactors proportional to $c$.
 From our arguments however, it is not obvious how the mutual information behaves at the transition point $n=n_c$, where the vertex operator is marginal. We will come back to this point in Sec.~\ref{sec:marginal}.
\subsection{Path-integral and replicas}
\label{sec:replicas}
For more general models or CFTs it is not so easy to use the explicit form of the action as in the Luttinger liquid. The standard method is to introduce $n$ replicas, with $n$ an integer greater than one. We have
\begin{equation}\label{eq:replicas}
 \mathcal{Z}^{(n)}=\frac{\mathcal{Z}_{\rm rep}}{\mathcal{Z}^n},
\end{equation}
where $\mathcal{Z}$ is the partition function of the infinite cylinder (strip). 
$\mathcal{Z}_{\rm rep}$ is the partition function of a replicated system: we have $n$ independent copies of the cut-cylinder (cut-strip) geometries shown in Fig.~\ref{fig:geometry}, but where the configurations of all copies are identified along the cut. Note that for the Shannon entropy of the full chain this geometry can be folded, and seen as a system of $2n$ semi-infinite cylinders (strips) glued along their common boundary.

It is also instructive to derive the free boson result before the transition ($n<n_c$), using replicas as in Ref.~[\onlinecite{FradkinMoore}]. We introduce $n$ copies $\varphi_i$, for $i=1,\ldots,n$ of the boson field, with total action
\begin{equation}
 {\bf A}=\sum_{i=1}^n \mathcal{A}_i\quad,\quad {\cal A}_i=\frac{g}{4\pi}\int_{0}^{L} dx_i \int_{-\infty}^{\infty} d\tau_i \left(\nabla \varphi_i\right)^2.
\end{equation}
All these fields are independent in the bulk, but they are stitched together at the slit. Neglecting the ambiguities in the compactification that do not matter at the leading order in this geometry (see the discussion after (\ref{eq:smp})), one can introduce the orthogonal transformation \cite{FradkinMoore} \mbox{$\tilde{\varphi}_1=\frac{1}{\sqrt{n}}\sum_i \varphi_i$} as well as $\tilde{\varphi}_i=\frac{1}{\sqrt{2}}\left(\varphi_{i}-\varphi_{i-1}\right)$  for $i=2,\ldots,n$. In terms of the new fields the action decouples in the bulk, ${\bf A}=\sum_i \tilde{\mathcal{A}}_i$. Since all the fields $\varphi_{i\geq 2}$ have to match on the slit, the new fields $\tilde{\varphi}_i$ vanish on it, hence they obey a Dirichlet boundary condition. The remaining field $\tilde{\varphi}_1$ fluctuates freely at the slit and cancels with one of the $n$ normalization partition functions $\mathcal{Z}$. In the end we recover Eq.~(\ref{eq:freebosonZreplicas}), and the result (\ref{eq:freeboson}) follows.

For the Ising CFT, such a gluing of CFT already becomes nontrivial. As can be seen in a Landau-Ginzburg (``$\phi^4$'') point of view (see e.g. Ref.~[\onlinecite{Oshikawa}]), the total bulk action does not decouple in terms of similar orthogonal transformation of the fields. Any unitary minimal model has a similar representation, and would suffer from this problem. However, previous numerical simulations of the R\'enyi entropy of the full chain \cite{SFMP,SMP2,Zaletel} have found results consistent with a boundary transition already at $n=n_c=1$. This implies that the entropy should be dominated by ordered configurations for any $n>1$, similar to what happens in the free boson case for $n>n_c$. Hence the copies  decouple, and we expect the RMI to be given by the simple formula
\begin{equation}\label{eq:rmi_conj}
 I_{n>1}(\ell,L)=\frac{c}{4}\frac{n}{n-1}\ln \left[\frac{L}{\pi}\sin \frac{\pi \ell}{L}\right].
\end{equation} 
Although it can be justified for large $n$ \cite{Cardy_bfield}, it is not obvious from perturbative RG arguments why this should happen already for $n>1$ \footnote{I thank John Cardy for pointing that out to me.}. The couplings near the slit are enhanced in the replica picture, or equivalently the temperature is lowered. One would naively expect these to be responsible for the ordering of the boundary.  
However, a typical scenario for a boundary in 2d classical critical systems with positive local degrees of freedom is that of the \emph{ordinary} transition, where a change of couplings near a boundary does not order it \cite{Diehl}. Here the gluing of $n$ copies appears to escape this scenario, and the slit on which the copies are stiched does  order. In surface critical phenomena language this is an \emph{extraordinary} transition \cite{Diehl}.

We have at present no analytical understanding of this observation. Similar gluing of 2d Ising models have already been studied in the literature (see e.g. \cite{IgloiTurbanBerche,Tsvelik1}), but in slightly different limits. Let us finally mention that universal scaling forms proportional to $n/(n-1)$ have been found in the 2d classical R\'enyi mutual information \cite{GMI} for Ising, as well as in the R\'enyi entropy of the 2d quantum transverse field Ising model \cite{AletShannon}. In both cases the underlying ordering assumption is easier to justify: for the former the critical system is coupled to a \emph{bulk} in the ordered phase \cite{Melkoinf3}, while for the latter the higher dimensionality makes an extraordinary transition more likely. 

To confirm our ordering assumption, we have computed numerically the spin-spin correlation function along the slit in the replicated geometry corresponding to $\mathcal{Z}^{(n)}$ ($n=1$ is the usual bulk spin-spin correlation function). For $n>1$ the data shows that this correlation becomes ordered, and the copies effectively decouple. We refer to Appendix~\ref{sec:correlations} for the details. Such a behavior is also strikingly different from the free boson case, where the boundary is still critical with a modified stiffness $g^\prime=ng$, see Eq.~(\ref{eq:changedstiffness}). We therefore still expect critical correlations for $n<n_c$ for the free boson, and ordered ones for $n>n_c$. This has already been shown \cite{Kumano} numerically in lattice discretizations such as the XXZ spin chain.

We present numerical results for the mutual information that support our conjecture (\ref{eq:rmi_conj}) in Sec.~\ref{sec:numerics}. As in the free boson case, our arguments do not predict the behavior of the mutual information exactly at the transition point ($n=1$ here). This study is deferred to Sec.~\ref{sec:transitions}, where we rely on numerics.
\section{Numerical checks of the CFT}
\label{sec:numerics}
In this section we perform various numerical checks of the conformal scaling in lattice discretizations of CFTs with central charge $c=1$ and $c=1/2$. 
\subsection{Lattice computations}
\label{sec:computations}
We focus here on two spin chains. The first is the antiferromagnetic XXZ chain
\begin{equation}
H=\sum_{i=1}^L \left(\sigma_i^x\sigma_{i+1}^x+\sigma_i^y\sigma_{i+1}^y+\Delta \sigma_i^z \sigma_{i+1}^z\right),
\end{equation}
which is (for $-1<\Delta\leq 1$) a Luttinger liquid CFT ($c=1$). Here the basis generated by the eigenstates of the $\sigma_j^z$ is most natural, as the bosonic field is diagonal in such a basis \cite{Giamarchi}. 
Even though the chain is integrable, it is in practice not easy to exploit the additional structure to compute the Shannon entropy (see however Refs.~[\onlinecite{Pozsgay,Caux}] for an explicit computation of $S_{\infty}(L,L)=-\ln p_{\uparrow\downarrow\ldots\uparrow\downarrow}$ in the periodic case). Here we generate the ground-state wave function using the Lanczos algorithm. For a periodic system one can use translational invariance, the reflection symmetry, and combine these with the $U(1)$ and particle-hole symmetry. Doing so system sizes up to $L=36$ can be reached with a reasonable amount of computer effort. 

The other is the quantum XY chain in transverse field:
\begin{equation}
 H=-\sum_i\left[ \left(\frac{1+\gamma}{2}\right)\sigma_i^x\sigma_{i+1}^x+\left(\frac{1-\gamma}{2}\right)\sigma_i^y\sigma_{i+1}^y+h\sigma_i^z\right]
\end{equation}
For $h=1$ and $\gamma>0$, its long distance behavior is described by the Ising $c=1/2$ CFT ($h=1$ and $\gamma=1$ is the Ising chain in transverse field --ICTF). Note that for $\gamma=0,h=0$ we recover the XXZ model at $\Delta=0$. The main advantage of this chain is that it can be written in terms of free fermions, and numerical computations are somewhat simplified. Indeed, performing a Jordan-Wigner transformation
\begin{eqnarray}\label{eq:jw1}
 \sigma_j^z&=&2c_j^\dag c_j-1,\\\label{eq:jw2}
\frac{\sigma_j^x+i\sigma_j^y}{2}&=& \exp\left(i\pi \sum_{l=1}^{j-1}c_l^\dag c_l\right)c_j^\dag,
\end{eqnarray}
allows to express $H$ as a quadratic form in the $c_i,c_i^\dag$, which may be diagonalized by a Bogoliubov transformation
\footnote{The Bogoliubov transformation is
$c_j^\dag=\sum_k (u_{kj} d_k^\dag+v_{kj}d_k)$, where the new fermions operators $d_k,d_k^\dag$ also obey the canonical anticommutation rules. In case of periodic boundary conditions $k$ also labels momentum.
}.
Each probability in subsystem $A$ can be obtained exactly, following e.g. Ref.~[\onlinecite{SMP2}]. We have
\begin{equation}\label{eq:probadet}
 p_{\mu}=\det_{1\leq i,j\leq \ell}\left(\,\left\langle f_i^\dag \left(f_j^\dag+f_j\right)\right\rangle_{\!L}\,\right),
\end{equation}
where $\langle \ldots\rangle_L$ denotes the average in the ground-state of the chain of size $L$, and $f_i^\dag=c_i^\dag$ (resp. $f_i^\dag=c_i$) if $\sigma_i^z=\uparrow$ (resp. $\sigma_i^z=\downarrow$). There are analogous formulae for systems $B$ and $A\cup B$, respectively as $(L-\ell)\times(L-\ell)$ and $L\times L$ determinants. Using this, the entropy in the $\sigma^z$ basis can be obtained by brute-force summation over the $2^\ell$ configurations, with total complexity $\ell^3 2^\ell$. The number of determinants to compute can be slightly reduced by making use of the lattice symmetries. Since each of the probabilities can be computed independently, parallelization is also trivial. In practice we can push the numerics up to $\ell \simeq 40$ independent on $L$, which significantly improves on Ref.~[\onlinecite{AlcarazRajabpour}], another advantage being that the data is exact up to machine-precision.

For the Ising chain the most natural local basis is generated by the $\sigma_j^x$, as these correspond to the actual spins in the classical two-dimensional model. In a periodic chain, it is also possible\cite{SFMP} to obtain the entropy in this basis, using the Kramers-Wannier duality $\sigma_j^z\to\tilde{\sigma}_{j-1}^x\tilde{\sigma}_j^x$, $\sigma_j^x\sigma_{j+1}^x\to \tilde{\sigma}_j^z$, which maps $H$ onto itself at the critical point. We get
\begin{equation}
 p(\sigma_1^x,\ldots,\sigma_L^x)=\frac{1}{2}p(\tilde{\sigma}_1^z,\ldots,\tilde{\sigma}_L^z),
\end{equation}
and evaluate $p(\tilde{\sigma}_1^z,\ldots,\tilde{\sigma}_L^z)$ using Eq.~(\ref{eq:probadet}). 
The factor $1/2$ is due to the fact that the mapping is two to one. The complexity for the subsystem entropy is slightly greater ($2^L L^3$) than in the $z$ basis, as one needs to generate all the probabilities of the full chain to access the subsystem (marginal) probabilities.
For open chains the Kramers-Wannier duality does not map $H$ onto itself, and this trick does not work anymore. The free fermions method does not outperform the Lanczos algorithm in this case.  
\subsection{Results for the free boson}
\begin{figure}[htbp]
 \includegraphics[width=8.5cm]{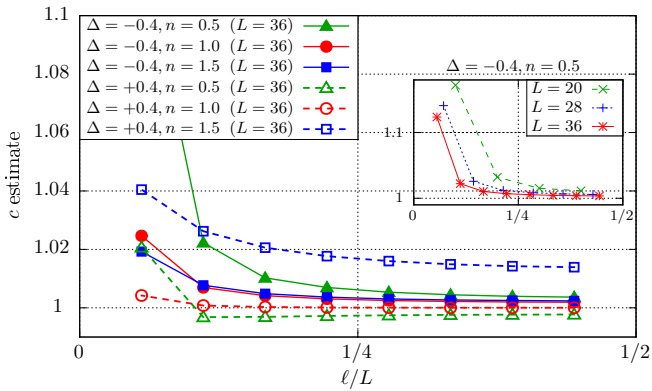}
 \caption{Numerical checks for the periodic $XXZ$ chain. We plot the $c$ estimate (\ref{eq:cestimate}) for two different values of the anisotropy ($\Delta=\pm 0.4$) and $3$ values of the R\'enyi index ($n=0.5,1,1.5$). The data shows excellent agreement with the conformal scaling and a central charge $c=1$. Inset: example of convergence of the $c$ estimate in the finite-size data.}
 \label{fig:LLperiodic}
\end{figure}

We first focus our attention on the periodic XXZ chain in the $z$ basis. In this chain the least irrelevant vertex operator allowed by the lattice symmetries is $V_2=\cos \left(\frac{2}{r}\varphi\right)$, so that (see Eq.~(\ref{eq:nc}))
\begin{equation}
 n_c=4K.
\end{equation}
The Luttinger parameter is known to be \cite{Giamarchi}
\begin{equation}
 K=\left(2-\frac{2}{\pi}\arccos \Delta\right)^{-1}.
\end{equation}
We test Eq.~(\ref{eq:freeboson}) for different values of $n$ and $\Delta$ before the transition ($n<n_c$), and plot the ratio
\begin{equation}\label{eq:cestimate}
 \frac{I_n(\ell,L)-I_n(L/2,L)}{\frac{1}{4}\ln \sin \frac{\pi \ell}{L}}
\end{equation}
in Fig.~\ref{fig:LLperiodic}. Provided (\ref{eq:freeboson}) is correct, this should give $c$ for all $\ell/L$. The results are, within at worst a few percents, compatible with this. Note that in principle the conformal limit is reached only when $\ell\gg 1$, so that we expect the result to deteriorate at small aspect ratios $\ell/L$. 
Finite-size effects are bigger when $\Delta<0$ and $n<1$, or $\Delta>0$ and $n>1$.
However, as illustrated in the inset, the accuracy improves as we increase the system size, so we expect a slow convergence towards $c=1$.
In practice the central charge estimate can be much improved by extrapolation, as is explained in Sec.~\ref{sec:transitions}. At the Shannon point such a procedure is not  needed, as the finite-size data already shows quasi-perfect agreement with the prediction. Remarkably, this holds even at the $SU(2)$ symmetric point $\Delta=1$, were a marginal bulk operator generates logarithmic corrections to correlation functions and free energies. Here this is not the case, and the ``raw'' data gives $c=1.007(10)$. Outside of the Shannon point, finite-size effects at $\Delta=1$ are substancial.  

When $n$ gets closer to $n_c$ (see blue square dashed curve in Fig.~\ref{fig:LLperiodic}) the agreement starts to deteriorate, as we approach the boundary KT transition discussed in Sec.~\ref{sec:freebosoncft}. For $n>n_c$, Eq.~(\ref{eq:freeboson_afternc}) should hold, and we also checked that this is the case. We differ the numerical results at the most interesting point $n=n_c$ to Sec.~\ref{sec:marginal}.
\subsection{Results for the Ising CFT}
\begin{figure}
 \includegraphics[width=8.5cm]{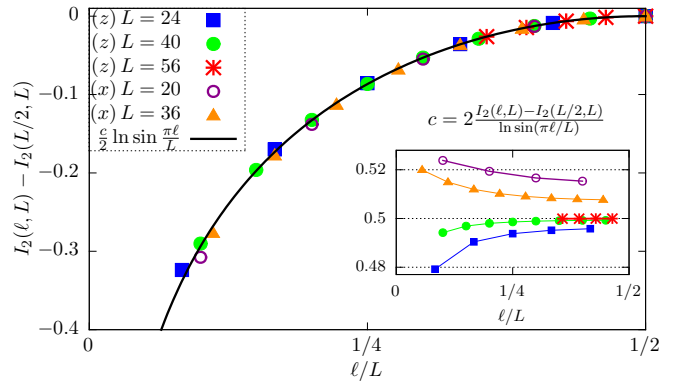}
 \caption{Second R\'enyi mutual information $I_2(\ell,L)-I_2(L/2,L)$ in the periodic Ising chain, and comparison with CFT (black curve). Inset: central charge extraction. Due to the huge computational costs involved, not all data is shown for the biggest system size $L=56$ in the $z-$basis.}
 \label{fig:Isingperiodic}
\end{figure}

As an example of $n>1$ simulation in the Ising universality class, we computed $I_2(\ell,L)$ in the Ising chain in transverse field (ICTF). The simulations were performed both in the $x-$ and $z-$ basis. Assuming the transition argument of Sec~\ref{sec:replicas}, the CFT predicts a scaling
\begin{equation}\label{eq:I2Ising}
 I_2(\ell,L)=\frac{c}{2}\ln \left(\frac{L}{\pi}\sin \frac{\pi \ell}{L}\right)+O(1).
\end{equation}
In Fig.~\ref{fig:Isingperiodic} we show $I_2(\ell,L)-I_2(L/2,L)$ for several system sizes, and compare it to the CFT result. As can be seen the agreement is excellent, and improves as $L$ gets larger. In the inset we also perform a central charge extraction similar to that done in Fig.~\ref{fig:LLperiodic}. We plot the ratio
\begin{equation}
 \frac{I_2(\ell,L)-I_2(L/2,L)}{\frac{1}{2}\ln \sin \frac{\pi \ell}{L}}
\end{equation}
as a function of $\ell/L$. The central charge extracted from this method gives an excellent agreement with $c=1/2$ in the $z-$basis. In the $x-$ basis there are slightly bigger finite-size effects, and the numerical data for our largest system size agrees up to $1.5$ percents. However the trend towards $c=1/2$ is clear.

All these numerical results nicely agree with our predictions, whichever the local basis ($x$ or $z$) we choose. This basis independence can be justified in the following way, assuming the phase transition scenario. For $n>1$ (universal terms in) the entropy will be dominated by the ordered configurations. In the $x$-basis these are $\ket{\uparrow\ldots \uparrow}_x$ and $\ket{\downarrow\ldots\downarrow}_x$. Since these correspond to the actual classical spin configuration in the 2d Ising model, the expected CIBC is the \emph{fixed} boundary condition. In the $z-$ basis only one configuration dominates, namely $\ket{\uparrow\ldots\uparrow}_z$. In the classical model this corresponds to a superposition of all possible spin configurations. Hence the appropriate CIBC is the \emph{free} boundary condition. However we have seen in Sec.~\ref{sec:efp} that the conformal scaling is insensitive to the precise nature of the (conformal) boundary conditions, and this justfies why (\ref{eq:I2Ising}) holds in both basis. We have also checked that the shape function is half the periodic one in open chains, as predicted by CFT.  

\section{Transitions and transition points}
\label{sec:transitions}
The aim of this section is to perform a detailed study of the transitions happening as a function of the R\'enyi parameter $n$ for the Ising and free boson universality class. In particular, we try and extract the universal prefactor $b_n$ with the highest accuracy possible. Recall we expect the mutual information to behave as
\begin{equation}\label{eq:RMIscaling2}
 I_n(\ell,L)=\frac{b_n}{4}\ln\left[\frac{L}{\pi}\sin\left(\frac{\pi \ell}{L}\right)\right].
\end{equation}
There are several different methods to extract $b_n$ from the numerical data. The main two we use are the following. Since the results of Figs.~\ref{fig:LLperiodic},\ref{fig:Isingperiodic} appear most accurate near $\ell=L/2$, we first look at a discrete derivative of the mutual information around this point
\begin{equation}\label{eq:deltadef}
 \delta_n(L)=\frac{2L^2}{\pi^2}\left[I_n(L/2,L)-I_n(L/2+2,L)\right]
\end{equation}
for large $L$. We consider only total system sizes $L$ multiple of four, so that all sizes are even, and potential parity effects are avoided. Assuming (\ref{eq:RMIscaling2}), $\delta_n(L)$ should scale as
\begin{equation}\label{eq:derivative}
 \delta_n(L)=b_n+o(L^0),
\end{equation}
and so can be used to extract $b_n$. In case the structure of the subleading corrections to (\ref{eq:derivative}) can be determined, extremely accurate values of $b_n$ can be extracted. For example including corrections of the form $\sum_{p=1}^3\alpha_p L^{-p}$ in (\ref{eq:derivative}) yields $b_{1/2}=c=1.001(2)$ for the worst data set ($n=1.5,\Delta=0.4$) in Fig.~\ref{fig:LLperiodic}. We call this method the ``discrete derivative method''.

An alternative procedure is to look at a finite subsystem of size $\ell$ in an infinite system ($L\to \infty$) \cite{PhDStephan}. The R\'enyi entropy is given in this case by
\begin{equation}\label{eq:infscaling}
 S_n(\ell,L\to \infty)=a_n\ell+\frac{b_n}{8}\ln \ell+\mathcal{O}(1),
\end{equation}
where $a_n$ is a line free energy. Numerical computations for free fermionic systems simulated in $z$-basis are possible, as is explained in Sec~\ref{sec:computations}. The logarithm is subleading, however the linear term can be substracted off by studying $S_n(\ell,\infty)-S_n(\ell,\ell)$, because $a_n$ does not depend on the precise geometry. As in the previous method, the results can be improved by extrapolation. In the following we will use both methods to extract $b_n$; their respective merits depend on the physical systems considered, and on our ability to guess the correct structure of the subleading corrections.
\subsection{Transitions}
Let us now demonstrate our results for the universal coefficient $b_n$. From the arguments presented above, we expect a phase transition in both the XX (\ref{sec:transition_xx}) and Ising (\ref{sec:transition_ising}) chains.
\subsubsection{XX chain}
\label{sec:transition_xx}
We start with the XX chain, as an example of the free boson described in Sec.~\ref{sec:freebosoncft}. Our arguments predict a transition at $n=n_c=4$. The expected result from CFT is
\begin{equation}\label{eq:xx_bn}
 b_n=\left\{
 \begin{array}{ccc}
  c&,&0<n<4\\\\
  \frac{n}{n-1}c&,&n>4
 \end{array}
 \right.
\end{equation}
with a central charge $c=1$.
\begin{figure}[htbp]
 \includegraphics[width=8.5cm]{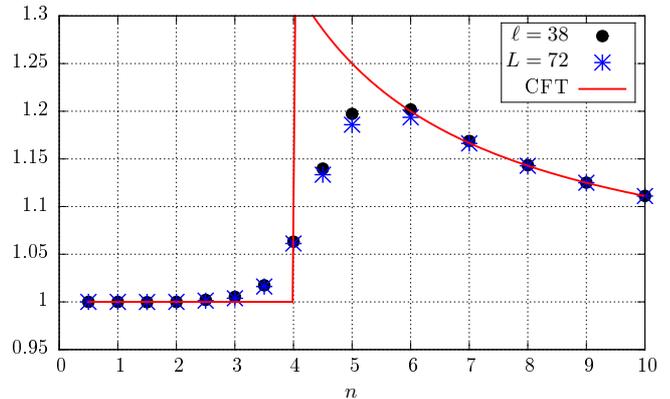}
\caption{Numerical extraction of $b_n$ in the XX chain. Black dots: extrapolation from the entropy in the infinite limit, up to subsystem sizes $\ell=38$. Blue stars: discrete derivative method, up to total system size $L=72$. The data is compared to the CFT prediciton of Eq.~(\ref{eq:xx_bn}) (thick red curve).}
 \label{fig:XX_transition}
\end{figure}
The discussion of the marginal point $n=4$ is differed to Sec.~\ref{sec:marginal}. The numerical results are shown in Fig.~\ref{fig:XX_transition}, where  
Black dots are the infinite method, and blue stars the derivative method. Both agree very well with the prediction, except close to the transition, where finite-size effects become substancial. This is expected, as the prediction has a discontinuity in the thermodynamic limit.  
Sufficiently far from the transition the agreement becomes excellent, e.g. both methods give $|b_n-1|<10^{-3}$ for $n=1/2,1,3/2,2$.
\subsubsection{Ising chain}
\label{sec:transition_ising}
\begin{figure}[htbp]
 \includegraphics[width=8.5cm]{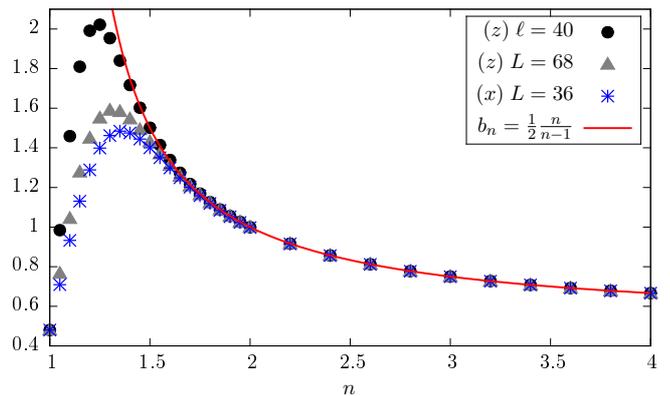}
 \caption{Numerical extraction of $b_n$ in the periodic ICTF. Black dots: extrapolation from the entropy in the infinite limit, up to subsystem sizes $\ell=40$, in the $z$ basis. Grey triangles: discrete derivative method, up to total system size $L=68$, $z$ basis. Blue stars: derivative method, up to $L=36$, $x$ basis. Red curve is the CFT result, assuming the phase transition argument.}
 \label{fig:transition}
\end{figure}

We now turn our attention to the ICTF, and perform the same extraction of $b_n$. As in the previous section, the entropy is studied both in the $z$-basis and $x$ basis (where the infinite method is not available). The numerical results are shown in Fig.~\ref{fig:transition}. 

We comment here on the results. For $n\gtrsim 1.7$ all three method agree extremely well with each other, and also with the CFT prediction for decoupling copies, $2b_n=n/(n-1)$. However, the discrete derivative method results deteriorate sooner than their infinite counterpart. This is probably a strong finite-size effect, as the trend is towards the curve as $L$ gets larger. Indeed, we use an extrapolation of the form $\alpha_1/L+\alpha_2/L^2+\alpha_3/L^3$ for all $n$, but we observed that the leading correction becomes slower than $L^{-1}$ when $n$ approaches one. The infinite method seems to suffer less from this problem. For it we extrapolate to $\beta_1(\ln \ell)/\ell+\beta_2(\ln \ell)/\ell^2$, where $\beta_p(x)$ is a polynomial of degree $p$ in $x$. Such corrections are inspired by the $n\to \infty$ limit studied in Sec.~\ref{sec:efp}, where they are known to appear\cite{logloverl}. We do not expect the derivative method to have such corrections: they vanish due to their antisymmetry with respect to $\ell\to L-\ell$ in this particular geometry \cite{emptiness}. 

These observations, together with the quasi-perfect crossings of the finite-size data at $n=1$, support the idea of a phase transition at $n=n_c=1$. Similar transitions were observed in the R\'enyi entropy of full chains \cite{SMP2,Zaletel}. 
\subsection{The transition points}
\label{sec:marginal}
The exact transition points $n=n_c$ are potentially the most interesting, and seem nontrivial from a boundary CFT perspective. It is not even guarantied that the conformal scaling survives. As we shall see the numerical results suggest that it does in \emph{periodic chains}, albeit with a nontrivial value of $b_{n_c}$.
\subsubsection{XX chain}
\begin{figure}[htbp]
 \includegraphics[width=8.5cm]{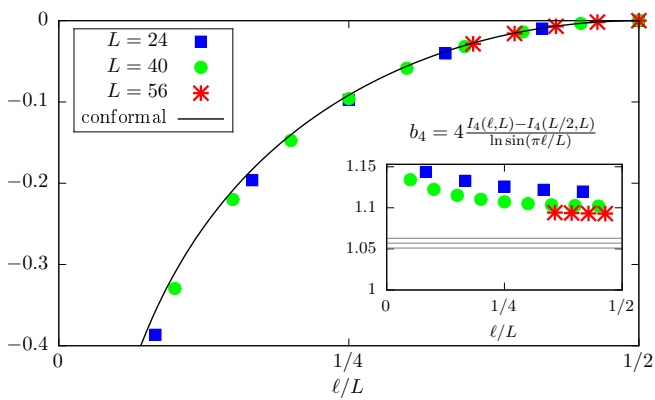}
 \caption{Fourth RMI $I_4(\ell,L)$ in the periodic XX chain and ``conformal'' curve with $b_4$ given by (\ref{eq:b4}). We show the data for $L=24,40,56$. Not all data points are shown for $L=56$ due to their huge computational costs. Inset: extraction of $b_4$, and comparison with (\ref{eq:b4}) including error bars.}
 \label{fig:XX_marginal}
\end{figure}

We first start with the XX chain. We study the R\'enyi mutual information for $n=n_c=4$. The numerical results for several system sizes are shown in Fig.~\ref{fig:XX_marginal}.
As can be seen the finite-size effects are somewhat bigger, but the data appears consistent with a conformal scaling. Using both the infinite \footnote{In this precise case the method can be slightly improved using the exact result for $S_4(L,L)$ derived in the appendix \ref{sec:exact}, that gives a leading term $a_4=\frac{2+\ln 2}{6}$ in Eq.~(\ref{eq:infscaling}).} 
and derivative extrapolation methods, we arrive at
\begin{equation}\label{eq:b4}
 b_4=1.057(6).
\end{equation}
The relatively big error bar is due to the fact that it is difficult to guess the precise form of the (rather slow) subleading corrections. This is the only instance in this paper were we did not manage to exclude a slight violation of the conformal scaling. Since $n=n_c$ corresponds here to a point where a cosine is marginal, this number is also possibly non-universal.
\subsubsection{Ising chain}
\begin{figure}[htbp]
 \includegraphics[width=8.5cm]{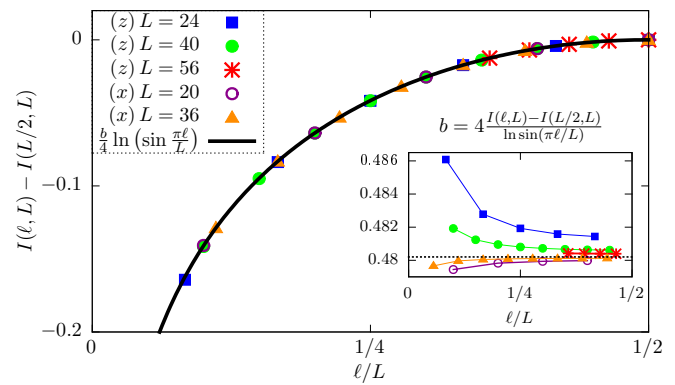}
 \caption{Shannon mutual information in the ICTF. The data for $L=20,36$ is shown in the $x$-basis, and $L=24,40,56$ in the $z$-basis. The conformal scaling is obeyed, with a nontrivial value of $b_1$, very close to $0.48$.}
 \label{fig:Ising_marginal}
\end{figure}

We now finally arrive at the Shannon $n=1$ point in the ICTF, where $b_1=c=1/2$ has been conjectured \cite{AlcarazRajabpour}. The numerical results in the $x$ and $z$ basis are shown in Fig.~\ref{fig:Ising_marginal}.
As can be seen, the data disagrees with the conjecture. The conformal scaling is almost perfectly obeyed by the finite-size data, but with prefactor $b_1\simeq 0.48$. This result is highly nontrivial, and we have no theoretical justification of it.

Contrary to the XX case at $n=4$, finite-size effects are quite small, and extremely accurate estimates for $b_1$ can be obtained by extrapolation. For example, the infinite method gives, with the power-series corrections $\sum_{p=0}^5 \alpha_p/\ell^p$, the estimate $b_1=0.480163(1)$. We did not observe any $\ell^{-1}\ln \ell$ terms at the Shannon point. However, and contrary to what happens in Sec~\ref{sec:transition_ising}, it is slightly outperformed by the discrete derivative method in the $z$ basis. The details of the fitting procedure are summarized in Tab.~\ref{tab:shannonfits}. Our best estimate is
\begin{equation}\label{eq:b_estimate}
 b_1=0.4801629(2).
\end{equation}
We remark that there is a compromise to make regarding the order at which we truncate the power-series correction. While going further is expected to improve the precision of the estimates, it also requires to keep more data points corresponding to smaller system sizes in the fit. Also, one needs to make sure that the numerical results from which we extrapolate are accurate enough: for the biggest system sizes ($\ell\simeq 40$) considered here, a correction of the form $1/\ell^6$ probes digits further than the tenth after the dot. Considering the exponential number of terms in the sum defining the entropy, these may start to be affected by rounding errors for the very largest system sizes.

 \begin{table*}
 \centering
\begin{tabular}{|c|c|c|c|||c|c|c|c|||c|c|c|}
 \hline
 $L$&$\delta_1^{(x)}(L)$&$b_1^{(x)}$, $q=2$&$b_1^{(x)}$, $q=4$&
 $L$&$\delta_1^{(z)}(L)$&$b_1^{(z)}$, $q=2$&$b_1^{(z)}$, $q=4$&
 $\ell$&$\Delta S_1^{(z)}(\ell)$&$b_1^{(z)}$, $q=3$\\
 \hline
 $8$&0.5376927020744&&&$44$&0.4821782481856&0.48014888&0.480162731&$24$&0.6513337781172&0.48016183\\
  \hline
 $12$&0.5031759924179&0.4755626&&$48$&0.4818548467208&0.48015347&0.480162796&$26$&0.6560852353300&0.48016213\\
  \hline
 $16$&0.4927141649930&0.4792632&&$52$&0.4816035919149&0.48015636&0.480162833&$28$&0.6604879499003&0.48016236\\
  \hline
 $20$&0.4880901551716&0.4798696&0.48000656&$56$&0.4814044890086&0.48015825&0.480162856&$30$&0.6645896372727&0.48016248\\
  \hline
 $24$&0.4856308346210&0.4800414&0.48014577&$60$&0.4812440267753&0.48015952&0.480162872&$32$&0.6684288433522&0.48016264\\
  \hline
 $28$&0.4841647379924&0.4801047&0.48015958&$64$&0.4811128065469&0.48016040&0.480162884&$34$&0.6720371533968&0.48016266\\
  \hline
 $32$&0.4832196088812&0.4801321&0.48016184&$68$&0.4810041253011&0.48016102&0.480162894&$36$&0.6754407733241&0.48016277\\
  \hline
 $36$&0.4825744159185&0.4801454&0.48016241&$72$&0.4809130976976&0.48016147&0.480162904&$38$&0.6786616859782&0.48016280\\
  \hline
 $40$&0.4821142385019&0.4801524&0.48016262&$76$&0.4808360946837&0.48016179&0.480162908&$40$&0.6817185124936&0.48016283\\
  \hline
\end{tabular}
\caption{Summary of the extrapolation results for $b_1$ in both $x$ and $z$ basis, using two methods. We first show the raw data for $\delta^{(x/z)}_1(L)$, defined by Eq.~(\ref{eq:deltadef}). This should converge to $b_1$ in the limit $L\to \infty$. We also extracted improved estimates, fitting the windowed data $[\delta_1(L),\delta_1(L-4),\ldots,\delta_1(L-4q+4)]$ to $b_1+\sum_{p=2}^q \alpha_p L^{-p}$ (we found that the term in $L^{-1}$ vanishes). The results of this procedure are shown for $q=2$ and $q=4$. In the $z$ basis one can also use the ``infinite'' method, by studying $\Delta S_1(\ell)=S_1^{(z)}(\ell,\infty)-S_1^{(z)}(\ell,\ell)$ and fitting the data $[\Delta S_1(\ell),\Delta S_1(\ell-2),\ldots,\Delta S_1(\ell-2q-4)]$ to $\frac{b_1}{8}\ln \ell+\sum_{p=0}^q \alpha_{p} \ell^{-p}$. Here the extracted value of $b_1$ is shown for $q=3$. All the numbers we obtain remain quite stable when changing $L,\ell$ or the size of the window, confirming the validity of our fitting ans\"atze. Not all digits are shown due to space limitations.}
\label{tab:shannonfits}
 \end{table*}
 
To confirm the universality of this number, we also checked that it appears for any $\gamma>0$ in the XY chain in transverse field. For $\gamma=0.5,1.5$ and other values, we are easily able to reproduce the first six digits in (\ref{eq:b_estimate}) using the same extrapolation methods. As additional evidence we also checked for the presence of $b_1$ in the dominant eigenvector of the transfer matrix of the square lattice Ising model. A brute force diagonalization up to $L=14$ already gives $b_1=0.481(3)$, consistent with the result of Ref.~[\onlinecite{LauGrassberger}]. All these observations combined provide strong evidence for the universality of $b_1$. 
\subsubsection{The peculiar case of open chains}
\label{sec:openchains}
It is also tempting to look at the shape dependence of the mutual information with open boundary conditions. In systems or R\'enyi indices where the CFT arguments of Sec~\ref{sec:CFT} do apply, the mutual information is expected to be half that of a periodic system. 

Our numerical results show that this is not true for Ising at $n=1$. The numerical simulations clearly suggest the following form for the Shannon entropy
\begin{equation}\label{eq:shannonopen_scaling}
 S_1(\ell,L)=a\ell+b^\prime\left(\ln \ell\right)^2+f(\ell/L)\ln \ell+O(1),
\end{equation}
which implies
\begin{equation}\label{eq:smi_open}
 I_1(\ell,L)=\left[f(\ell/L)+f(1-\ell/L)\right]\ln \ell+O(1).
\end{equation}
This unusual scaling, with a shape-dependent logarithmic term, is very different from what we have seen before, and defies our simple CFT arguments. In practice, it is difficult to extract the shape-dependent $\ln \ell$ term in the mutual information. This is because an extrapolation with fixed aspect ratio $\ell/L$ is required, and due to commensurability effects this may require unreachable system sizes. However, we are able to extract $f(\ell/L)$ from the Shannon subsystem entropy in the $z$ basis, where slightly bigger system sizes are available. In practice we extrapolate $S_1(\ell,L)$ to $a\ell+b^\prime (\ln \ell)^2+\beta_0\ln \ell+\alpha_0+\beta_1\ell^{-1}\ln \ell+\alpha_1\ell^{-1}$ for fixed aspect ratio $\ell/L$. The $\ell^{-1}\ln \ell$ is a natural correction here, as can be seen by making the substitution $\ell\to \ell+\epsilon$ for some ultraviolet cutoff $\epsilon$, and expanding again in $\ell$. Using this method, we found very stable values of $f(\ell/L)$, which is good evidence for the correctness of our scaling ansatz.  

\begin{figure}[htbp]
 \includegraphics[width=8.5cm]{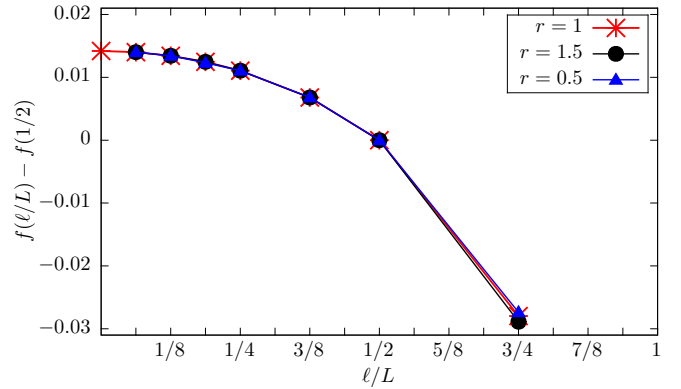}
 \caption{Numerical extraction of the shape-dependent logarithmic prefactor in the XY chain on the Ising critical line. We use system sizes up to $\ell=36$ for the fits. $\gamma=1$ (ICTF) as well as $\gamma=0.5$ and $\gamma=1.5$ are shown. The results  clearly support the universality of this shape-function.}
 \label{fig:I1Ising_open}
\end{figure}

The results are shown in Fig.~\ref{fig:I1Ising_open} for three different values of $\gamma$ in the XY chain, and strongly suggest that the shape function $f(\ell/L)-f(\ell/L=1/2)$ is universal. 
Note that due to the aforementioned commensurability effects\footnote{Indeed, for a fixed aspect ratio $\ell/L=p/q$, we need to study even system sizes also multiple of $p$.}, we have only access to a few particular aspect ratios.

The squared logarithmic divergence in (\ref{eq:shannonopen_scaling}) is also very intriguing. We checked that its value is independent on $\gamma$ in the XY chain, so that it is likely to be universal. We also checked that its coefficient is doubled for $S_1(L,L)$. This suggests that the corners with angle $\pi/2$ in the space-time geometry (see Fig.\ref{fig:geometry}(b)) are responsible for this enhanced scaling in the replica $n\to 1$ limit. Indeed there are twice as many such corners in $S_1(L,L)$ compared to $S_1(\ell<L,L)$, and none in a periodic system. Extracting $b^\prime$ accurately is more difficult than $b$ in the periodic geometry, as we have to take into account more corrections. This can nevertheless be done in the $z$ basis, where we pushed the numerics up to $\ell=40$ for $S(\ell,L\to \infty)$, which gives $b^\prime$, and up to $L=42$ for $S(L,L)$, which gives $2b^\prime$. Combining these two estimates we obtain
\begin{equation}
b^\prime=-0.02934(5). 
\end{equation}
In the $x$ basis smaller system sizes are available, and we cannot reach such a good precision. However the results appear consistent, within less than a percent, with the guess
\begin{equation}
 b^\prime_{(x)}=-b^\prime_{(z)}.
\end{equation}
This finding for $b^\prime_{(x)}$ is also consistent with the slow divergence of the $\ln L$ prefactor mentioned in the supplementary material of Ref.~[\onlinecite{Zaletel}].
We conjecture that these squared logarithms are a generic feature of spin chains described by minimal model CFTs. Such terms should also impact related quantities in classical systems. For example, the Shannon mutual information of a infinite cylinder (strip) in a 2d classical system is exactly (twice) the Shannon entropy of the dominant eigenvector of the corresponding transfer matrix \cite{GMI}. Hence by universality the $(\ln L)^2$ contribution should also appear in the 2d MI, most probably also in a finite system. It would be interesting to check that this is the case. 

Note however that these have not been observed for Luttinger liquids at $n=n_c$, where numerics\cite{SMP3} indicate a simple logarithmic term with a nontrivial prefactor. Interestingly, we are even able to prove this in the XX chain at half-filling. Indeed, using a result in Appendix \ref{sec:exact}, the fourth R\'enyi entropy of the open XX chain is exactly given, at half-filling, by
\begin{equation}
 S_4(L,L)=-\frac{1}{3}\ln\left[\frac{2^{L/2}}{(L+1)^L}\left(\frac{\Gamma\left[\frac{L}{2}+\frac{3}{4}\right]}{\Gamma\left[\frac{3}{4}\right]}\right)^2\right],
\end{equation}
where $\Gamma$ is the Euler Gamma function. An asymptotic expansion, using Stirlings formula, yields
\begin{equation}
 S_4(L,L)=\left(\frac{1}{3}+\frac{\ln 2}{6}\right)L-\frac{1}{6}\ln L+O(1).
\end{equation}
Most probably, the behavior of the entropy at the transition points in the two universality classes have different physical origins. 
\section{Conclusion}
\label{sec:Conclusion}
We have studied in this paper the R\'enyi and Shannon mutual information of one-dimensional quantum critical systems, using a combination of CFT and numerical techniques. Our main result is that the RMI for $n\geq 1$ follows the simple conformal scaling formula
\begin{equation}
 I_n(\ell,L)=\frac{b_n}{4}\ln \left[\frac{L}{\pi}\sin \left(\frac{\pi \ell}{L}\right)\right],
\end{equation}
for a periodic system cut into two subsystems of sizes $\ell$ and $L-\ell$. We have derived this result from CFT in the case of the free boson, and shown that $b_n$ is proportional to the central charge, except at a special point $n=n_c$ where it is unknown. At $n=1$ our analytical formula recovers the numerical result of Ref.~[\onlinecite{AlcarazRajabpour}]. 

The Ising universality class proved more tricky, as we were unable to formally derive the conformal scaling. However, inspired by the phase transitions observed in \cite{SMP2,Zaletel}, we showed numerically that the system orders at the slit in a replica picture for $n>1$. This implies the conformal scaling, and the result $b_n=c\frac{n}{n-1}$, independent of the local ($x$ or $z$) spin basis. We checked this prediction numerically, and found very good agreement. By construction this argument breaks down when $n\to1$. However the conformal scaling survives, with a value for $b_1$ that we computed numerically (see Eq.~(\ref{eq:b_estimate})). The Shannon point proves even more mysterious in open chains, where the full Shannon entropy has a $\left(\ln L\right)^2$ contribution, and the SMI diverges logarithmically with a shape-dependent prefactor (\ref{eq:smi_open}). We have currently no analytical understanding of these observations, and it would be highly desirable to make progress on this vexing problem.

The effect of having a R\'enyi index $n\neq 1$ is intuitively similar to a change of temperature in the 2d classical model, as can be seen from the difference between the  XXZ/six-vertex and the Ising results. Indeed, the former models possess a line of critical points, while the latter have a critical point that separates two gapped phases. All the competing orders are present in the (critical) ground-state wavefunction, and changing $n$ allows to distinguich between them. Such an interpretation should not be taken too litteraly, as is discussed in Sec.~\ref{sec:replicas}. It would even give a wrong $n_c$ for the Luttinger liquid, by neglecting the fact that the transition is a \emph{boundary} transition, not a bulk transition. Note however that $n$ is a true inverse temperature if we interpret the $p_i$ as Boltzmann weights for a 1d classical chain. In this case transitions at finite temperature are possible because the interactions are long-range (see Appendix.~\ref{sec:connection} for an illustration in the XX chain). Finally and due to the correspondence with the R\'enyi entanglement entropy of certain Rokhsar-Kivelson states \cite{SFMP}, these transitions also give us a lot of information about the entanglement spectrum in such states \cite{SMP2,Sondhispectrum}.

From our results for the Ising universality class it is tempting to conjecture a similar behavior for all minimal models (or any subset that closes under fusion rules). For example, Ref.~[\onlinecite{AlcarazRajabpour}] found $b_1 \simeq 0.79$ in the $3$-state quantum Potts model ($c=4/5=0.8$), and our findings indicate that this probably differs slightly from the central charge. Interestingly all these numbers should be universal: even though their exact value is not known exactly, they can still be used to identify the universality class. The fact that they are so close to the actual central charge is also very intriguing. Another interesting project would be to study the possible transitions as a function of $n$. For more complicated models the set of allowed conformal boundary conditions becomes larger, and one could imagine several transitions in the $b_n$ exponent or the R\'enyi entropy, with the conformal scaling still holding.

Another interesting direction would be to study the R\'enyi entropy when $n<1$ for non free bosonic theories. For such values the replica approach brings some results for the entropy of the full chain. For example $Z_{1/2}$ becomes the partition function of a single half-sheet and boundary CFT applies \cite{SMP2}. This is not the case anymore for the subsystem entropy, and one needs to rely on other methods. Even the numerical results show big finite-size effects, and it is difficult to distinguish between a conformal scaling with strong subleading corrections and a slightly enhanced logarithmic scaling\footnote{with exotic terms such as e.g. $\left(\ln \ln L\right)\times \ln L$.}.
\acknowledgments
I wish to thank Fabien Alet, Bertrand Berche, J\'er\^ome Dubail, Paul Fendley, Gr\'egoire Misguich and Vincent Pasquier for several stimulating discussions. I especially thank John Cardy for sharing some of his notes on the subject with me.
 The simulations were performed on the UVACSE ITS cluster. This work was supported by the US National Science Foundation under the grant DMR/MPS1006549.
\appendix
\section{Conformal mappings}
\label{sec:mappings}
In this appendix we recall the derivation of the conformal scaling of the emptiness formation probability in the periodic and open geometries (see Fig.~\ref{fig:geometry} (a) and (b)). We follow the method of Ref.~[\onlinecite{CardyPeschel}].
\subsection{Cylinder with a slit}
We start with the cylinder with a slit, relevant to periodic systems. 
\begin{figure}[htbp]\includegraphics{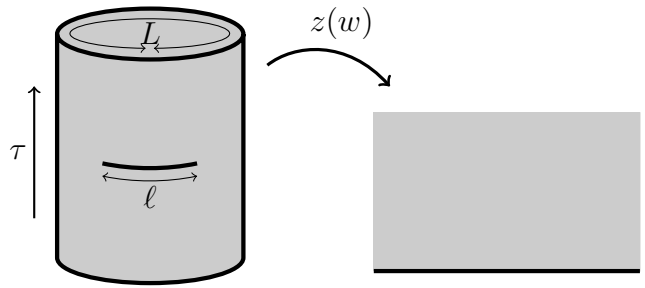}
 \caption{Conformal mapping from the infinite cylinder with a slit to the upper-half plane.}
 \label{fig:cyl_mapping}
\end{figure}
 To proceed we need the determine the average value of the stress-tensor\cite{Yellowpages}. It can be determined from the transformation law
 \begin{equation}
  T(w)=\left(\frac{dz}{dw}\right)^2T(z)+\frac{c}{12}\{z(w),w\}
 \end{equation}
were $\{z(w),w\}=z'''/z-3/2(z''/z')^2$ denotes the Schwarzian derivative. Using an inverse mapping of our geometry to the upper-half plane $\mathbb{H}=\{z,\; {\rm Im}\, z>0\}$, where $\braket{T(z)}=0$ by translational invariance, 
 we get
 \begin{equation}
  \braket{T(w)}=\frac{c}{12}\{z(w),w\}
 \end{equation}
In the slit-cylinder geometry an appropriate conformal mapping is given by
\begin{equation}
 z(w)=\sqrt{\frac{\sin \frac{\pi}{2L}\left(\ell+2w\right)}{\sin \frac{\pi}{2L}\left(\ell-2w\right)}},
\end{equation}
so that the expectation value of the stress tensor is
\begin{equation}
 \braket{T(w)}_{(a)}=\frac{\pi^2 c}{24L^2}\left(4+3\left[\frac{\sin \frac{\pi\ell}{L}}{\cos\frac{\pi\ell}{L}-\cos\frac{2\pi w}{L}}\right]^2\right)
\end{equation}
The variation of the free energy is given by \cite{CardyPeschel}
\begin{equation}
 -\frac{\delta \ln \mathcal{Z}_{\rm slit}}{\delta \ell}=\frac{1}{2\pi}\int_{\cal C}\braket{T(w)}dw 
\end{equation}
where the integral is taken over a contour ${\cal C}$ that encircles the slit. Using the residue theorem, we get
\begin{equation}
 -\frac{\delta \ln \mathcal{Z}_{\rm slit}}{\delta \ell}=\frac{\pi c}{8L}\cot \frac{\pi \ell}{L}
\end{equation}
which after integration $\mathcal{F}=\int_{a}^{\ell}\frac{\delta \mathcal{F}}{\delta \ell^\prime}d\ell^\prime$ ($a$ is a UV cutoff of the order of a lattice spacing), yields
\begin{equation}\label{eq:confscalingefp}
 -\ln \mathcal{Z}_{\rm slit}=\frac{c}{8}\ln \left(\frac{L}{\pi}\sin \frac{\pi \ell}{L}\right)+\ldots.
\end{equation}
The ellipsis stands for subleading terms. 
The leading $\ln L$ term in (\ref{eq:confscalingefp}) is a particular case of the Cardy-Peschel formula\cite{CardyPeschel}. For each corner with angle $\theta$, there is a contribution
\begin{equation}
 \left[\frac{c}{24}\left(\frac{\theta}{\pi}-\frac{\pi}{\theta}\right)+h\frac{\pi}{\theta}\right]\ln L,
\end{equation}
where $c$ is the central charge and $h$ the dimension of the possible boundary changing operator at the corner. 
\subsection{Strip with a slit}
\begin{figure}[htbp]
 \includegraphics{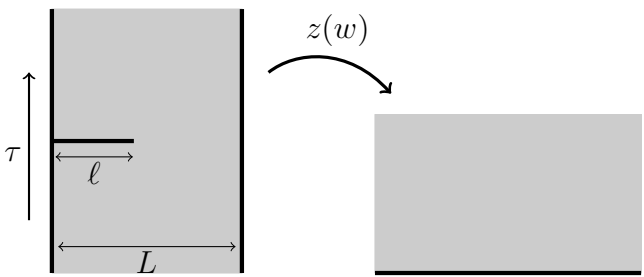}
 \caption{Conformal mapping from the infinite strip with a slit to the upper half plane. In our convention the slit is at $0\leq {\rm Re}(w)\leq \ell$ and ${\rm Im}(w)=0$.}
 \label{fig:strip_mapping}
\end{figure}
The open geometry is similar. Here the conformal mapping reads
\begin{equation}
 z(w)=\sqrt{1-\frac{\tan^2\frac{\pi w}{2L}}{\tan^2\frac{\pi \ell}{2L}}}
\end{equation}
and we have
\begin{eqnarray}\nonumber
 \braket{T(w)}_{(b)}&=&\frac{\pi^2 c}{96L^2}\left(4+3\frac{\sin^2 \frac{\pi \ell}{2L}}{\sin^2\frac{\pi w}{2L}}\left[\frac{3}{\cos \frac{\pi \ell}{L}-\cos\frac{\pi w}{L}}\right.\right.\\
 &+&\left.\left.\frac{1-\cos \frac{\pi \ell}{L}\cos \frac{\pi w}{L}}{\left(\cos \frac{\pi \ell}{L}-\cos \frac{\pi w}{L}\right)^2}\right]\right).
\end{eqnarray}
The partition function can be written as (see e.g. Ref.~[\onlinecite{Yellowpages}])
\begin{equation}
 \mathcal{Z}_{\rm slit}=\mathcal{Z}_{\rm slit}^{(0)}\braket{\phi(w_1) \phi(w_2)},
\end{equation}
where $w_1=i0^+$ and $w_2=i0^-$. The correlator can be obtained by a mapping to the upper half-plane, using the transformation law of the operator $\phi$, which is primary\cite{Yellowpages}. The respective images of $w_1$ and $w_2$ are $z_1=-1$ and $z_2=1$.
We get
\begin{equation}
 \braket{\phi(w_1)\phi(w_2)}=\left|\frac{L}{\pi}\tan \frac{\pi \ell}{2L}\right|^{-4h},
\end{equation}
and
\begin{equation}
 - \frac{\delta\ln \mathcal{Z}_{\rm slit}^{(0)}}{\delta \ell}=\frac{\pi c}{16L}\times \frac{-2+\cos \frac{\pi \ell}{L}}{\sin \frac{\pi \ell}{L}}.
\end{equation}
After integrating and combining the two contribution we finally obtain
\begin{equation}
 -\ln \mathcal{Z}_{\rm slit}=\frac{c}{16}\ln \left(\frac{\pi}{4L}\frac{\sin \frac{\pi \ell}{L}}{\tan^2\frac{\pi \ell}{2L}}\right)+4h\ln \left(\frac{L}{\pi}\tan\frac{\pi \ell}{2L}\right)+\ldots
\end{equation}
\section{Correlation functions for Ising in the replica systems}
\label{sec:correlations}
The aim of this appendix is to show some numerical results for spin-spin correlations at the slit in a replica picture. We focus on the Ising case, and show that the correlations are ordered when $n>1$. For convenience we consider the periodic case with $\ell=L$, relevant to the total R\'enyi entropy $S_n(L,L)$. However the ordering can also be observed for any $\ell/L$. Since $\ell=L$ we can fold the system, so that we are studying correlations at the binding of a $2n$-sheeted ``book'', where each of the sheets are independent except at the binding. To compute them we use the hamiltonian limit. The ground-state can be written, in the spin basis, as
\begin{equation}
\ket{\psi}=\sum_{\sigma} \psi_\sigma \ket{\sigma} 
\end{equation}
The correlation we look at is 
\begin{equation}
 C_n(\ell,L)=\Braket{\sigma_0^x\sigma_\ell^x}_{(n)}, 
\end{equation}
where $\Braket{\ldots}_{(n)}$ denotes the average in the ``r\'enyified'' ground state
\begin{equation}
 \Ket{\psi}_{(n)}=\frac{1}{\sqrt{Z_n}}\sum_{\sigma}\left(\psi_\sigma\right)^n \ket{\sigma}.
\end{equation}
Recall that the $\sigma^x$ correspond to the classical Ising spins in the 2d model, so we use this basis for the numerical computations.

For critical correlations conformal invariance implies
\begin{equation}\label{eq:corr_conf}
 C_n(\ell,L)\sim \left|\frac{L}{\pi}\sin \frac{\pi \ell}{L}\right|^{-\alpha_n},
\end{equation}
where $\alpha_n$ is the critical exponent. There are three values of $n$ for which the exponent can easily be determined:
\begin{itemize}
 \item The book with one sheet ($n=1/2$). The correlation is that of spins living at the boundary of a semi-infinite cylinder with free boundary condition. In this case the exponent is $\alpha_{1/2}=1$.
 \item The book with two sheets ($n=1$). This corresponds to the usual ground-state correlations, with the well known Onsager exponent $\alpha_1=1/4$.
 \item The book with an infinite number of sheets ($n\to\infty$). Here the r\'enyified ground-state is dominated by the two ordered states, and $\alpha_\infty=0$.
 \end{itemize}
\begin{figure}[htbp]
 \includegraphics[width=8.5cm]{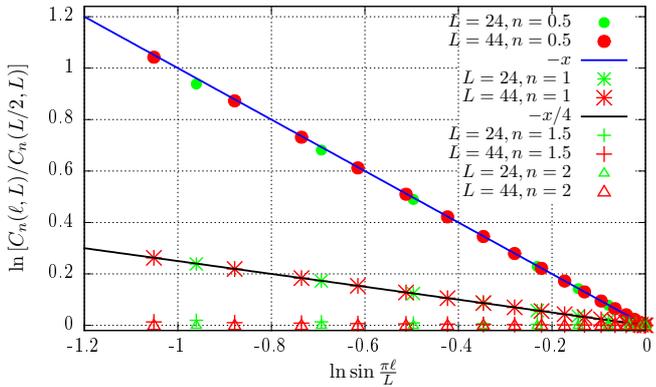}
 \caption{Extraction of the $\alpha_n$ exponent of Eq. (\ref{eq:corr_conf}), for $n=0.5,1,1.5,2$. The data for $n=0.5$ and $n=1$ agrees very well with the known exponents $\alpha_{1/2}=1$ and $\alpha_1=1/4$. The data for $n=1.5,2$ is consistent with an exponent zero, and therefore an ordering of the spins, at the binding of the book.}
 \label{fig:corr_extr}
\end{figure}
We extract the exponent numerically in Fig.~\ref{fig:corr_extr}, by plotting $\ln [C_n(\ell,L)/ C_n(L/2,L)]$ as a function of $\ln \sin \frac{\pi \ell}{L}$. As can be seen, the two first simple limits are recovered with good precision. The data for $n>1$ supports the idea of a phase transition, with an extracted $\alpha_n$ exponent already very close to $0$ for small system sizes; e.g. for $\ell/L=1/4$ the data points in Fig.~\ref{fig:corr_extr} are $\alpha_{0.5}=0.98812$, $\alpha_1=0.24923$, $\alpha_{1.5}=0.01831$, $\alpha_2=0.00084$ for total system size $L=24$, and $\alpha_{0.5}=0.99637$, $\alpha_1=0.24977$, $\alpha_{1.5}=0.01073$, $\alpha_2=0.00026$ for $L=44$. 
 This behavior is in sharp contrast with that of the Luttinger liquid, where the expected exponent from the replica picture is $\alpha_n=2K/n$. We show an explicit computation of this exponent for the XX chain at $n=2$ in Appendix.~\ref{sec:exact}.

\begin{figure}[htbp]
  \includegraphics[width=8.5cm]{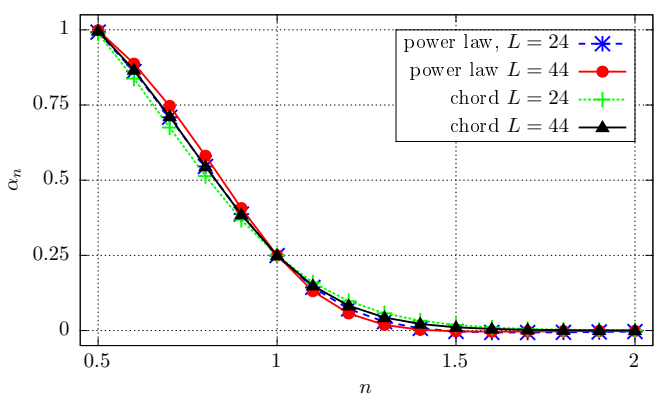}
  \caption{Spin-spin correlation exponent in the ``R\'enyified'' ground state $\ket{\psi}_{(n)}$, as a function of $n$. We show the finite-size exponent extracted from the two methods exposed in the text, and for system sizes $L=24$ and $L=44$.}
  \label{fig:exponent}
 \end{figure}

 In the region $1/2\leq n<1$ finite-size effects are very strong, and it is unclear what the exponent is. To illustrate this, we show in Fig.~\ref{fig:exponent} an extraction of the exponent using two slightly different methods. 
 The first to look at the ratio $\frac{C_n(L/4,L)}{C_n(L/2,L)}$, which should converge to $2^{\alpha_n/2}$ for large $n$. The second is to fit $C_n(L/2,L)$ to a power law $aL^{-\alpha_n}$. As can be seen in the figure, $\alpha_n$ varies significantly as the system size increases. The most likely scenario is be a slow convergence towards a constant $\alpha_n$ in this region, with a discontinuity at $n=1$. Such a picture would also be compatible with numerical results for subleading terms in the entropy of the full system\cite{SMP2,Zaletel}.

\section{Some exact results for the R\'enyi entropy of the full $XX$ chain}
\label{sec:exact}
This last appendix is devoted to the study of the R\'enyi entropy $S_n(L,L)$ of the full XX chain. In this simple model and geometry the R\'enyi entropy is related to a partition function for a 2d gas of particles on a ring, a discrete analog of the Dyson gas \cite{Dyson}. We use this connection to derive several exact results for integer $n$ R\'enyi indices. These are in agreement with the arguments presented in the main text, and offer supplementary evidence to the phase transition scenario for the free boson CFT.

The section is organized as follows. We start in \ref{sec:identities} by presenting some determinant and combinatorial identities that will be useful throughout. The connection between the entropy and the discrete gas is explained in \ref{sec:connection}, and a summary of the results in given in \ref{sec:dgresults}. Some of the technical details of the computations are finally gathered in \ref{sec:dgdetails}.
\subsection{Determinants and constant term identities}
\label{sec:identities}
\subsubsection{Vandermonde determinants}
Here we start by presenting some useful determinant identities. The first is the Vandermonde determinant, given by
\begin{eqnarray}\label{eq:vand}
V(u_1,u_2,\ldots,u_N)&=& \det_{1\leq j,k\leq N} \left(u_j^{k-1}\right)\\
&=&\prod_{1\leq j<k\leq N} (u_j-u_k).
\end{eqnarray}
A similar determinant, sometimes called symplectic Vandermonde, describes the ground-state of the open XX chain. 
\begin{eqnarray}\label{eq:svand}
 W(u_1,\ldots ,u_N)&=& \det_{1\leq j,k\leq N}\left(u_j^k-u_j^{-k}\right)\\\nonumber
 &=&\prod_{j=1}^N u_j^{-N}(1-u_j^2)\prod_{k>j}(u_j-u_k)(1-u_ju_k)
\end{eqnarray}
Certain powers of $V$ and $W$ can also be expressed as determinants. For example the fourth power of $V$ is related to a $2N\times 2N$ determinant:
\begin{equation}\label{eq:cvand}
 V(x_1,\ldots,x_N)^4=\det_{\begin{array}{c} \scriptstyle 1\leq j\leq N \\  \scriptstyle 1\leq k\leq 2N\end{array}} \left(\begin{array}{c}
u_j^{k-1}\\(k-1)u_j^{k-2}
\end{array}
\right).
\end{equation}
(\ref{eq:cvand}) can obtained by considering the limit
\begin{equation}
 \lim_{v_1\to u_1,\ldots,v_N\to u_N} \;\frac{V(u_1,v_1,\ldots,u_N,v_N)}{(u_1-v_1)\ldots(u_N-v_N)},
\end{equation}
and performing elementary row manipulation that leave the determinant invariant. A similar trick can be used on $W$. We get
\begin{equation}\label{eq:csvand}
 \det_{\begin{array}{c} \scriptstyle 1\leq j\leq N \\  \scriptstyle 1\leq k\leq 2N\end{array}} \left(\begin{array}{c}
u_j^{k}-u_j^{-k}\\k u_j^{k-1}+k u_j^{-k-1}
\end{array}
\right)=\frac{W(u_1,\ldots,u_N)^4}{\prod_{j=1}^{N}(u_j^2-1)}.
\end{equation}

\subsubsection{Dyson and Macdonald constant term identities}
\label{sec:dysonmacdonald}
Many of the partition functions shown in this appendix will follow from a constant term identity due to MacDonald \cite{Macdonald}:
\begin{equation}\label{eq:macdonald}
 {\rm CT} \left[\prod_{\alpha\in R}\left(1-e^\alpha\right)^k\right]=\prod_{i=1}^{N} \left(\begin{array}{c}kd_i \\k\end{array}\right)
\end{equation}
where the product runs over vectors of $\mathbb{R}^N$, elements of a crystallographic root system. ${\rm CT}$ stands for the constant term in the expansion of the product. The $d_i$ are a set of integers characteristic of the root system\cite{Macdonald}. 

The simplest (infinite) family of root systems is given by the series
\begin{equation}
 A_{N-1}=\{\pm\left(t_i-t_j\right)\;,1\leq i< j\leq N\},
\end{equation}
for which $d_i=i$ and (\ref{eq:macdonald}) reduces to
 the better known Dyson constant term identity \cite{Dyson,Good}
\begin{equation}
 {\rm CT} \left[\prod_{j\neq i}^{N} \left(1-e^{t_i-t_j}\right)^{n}.
 \right]=\frac{(nN)!}{n!^{N}}.
\end{equation}
In the following, we will also get the $C_N$ root systems, given by 
\begin{equation}
C_N=\{\pm \left(t_i\pm t_j\right),1\leq i<j\leq N\}\cup\{\pm 2t_i,1\leq i\leq N\}.  
\end{equation}
In this case, (\ref{eq:macdonald}) holds with
\begin{equation}\label{eq:di}
d_i=2i. 
\end{equation}
\subsection{R\'enyi-Shannon entropy and Dyson-Gaudin gas}
\label{sec:connection}
 We now establish the relation between the entropy and the thermodynamics of a discrete log-gas. The XX Hamiltonian we consider is given by
 \begin{equation}
  H=\sum_{j=1}^{L-1}\left(\sigma_j^x\sigma_{j+1}^x+\sigma_j^y\sigma_{j+1}^y\right)+H_{bound},
 \end{equation}
where $H_{bound}$ encodes the boundary conditions. In the following we wish to study both periodic boundary conditions ($H_{bound}=\sigma_L^x\sigma_1^x+\sigma_L^y\sigma_1^y$) and open boundary conditions
 ($H_{bound}=0$). In the latter case the relevant external conformal boundary condition is Dirichlet. To illustrate the effect of boundary changing operators, we also considered a case where the external boundary condition is Neuman. This can be achieved by applying any finite magnetic field along $x$ at the two boundary spins\cite{Affleck}. Here we focus on $H_{bound}=\sqrt{2}(\sigma_1^x+\sigma_L^x)$: the amplitude $\sqrt{2}$ (instead of a general $h$) is chosen because the diagonalization of $H$ simplifies considerably at this special point\cite{BilsteinWehefritz,Wehefritz}. An important difference with the first two cases is that the total magnetization $M=\sum_{j=1}^L \sigma_j^z$ is not conserved anymore.
 \begin{figure*}[htbp]
\includegraphics{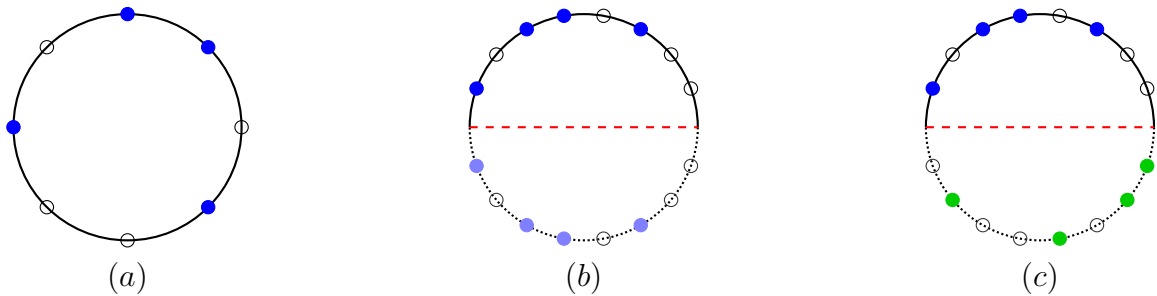}
 \label{fig:dysongaudin}
 \caption{Graphical representation of the Dyson-Gaudin gas corresponding to different boundary conditions. 
 The spin configuration $\ket{\uparrow\downarrow\uparrow\uparrow\downarrow\uparrow\downarrow\downarrow}$ with $L=8$ is shown. ($a$): Periodic case, the particles (up spins) are in blue. ($b$) Open case: each blue particle in the upper part interacts with the others and their mirror images (in lighter blue). ($c$) Open Neuman case: the particles are the originial ones, in blue, or the mirror images of the holes, in green. All of them interact with each other.}
\end{figure*}

 As is well known, such Hamiltonians can be mapped, via a Jordan-Wigner transformation (\ref{eq:jw1},\ref{eq:jw2}), onto a system of free fermions. Let us focus on periodic and open systems for now. The fermions label the positions of the up spins. Magnetization is conserved so there are a certain number, say $N$, of them. In position space, the wave function reads
 \begin{equation}\label{eq:wavedet}
  \psi(x_1,x_2,\ldots,x_N)=\det \left(\braket{c_{x_j}d_k^\dag}\right)_{1\leq j,k\leq N}
 \end{equation}
due to Wick's theorem. The expectation value is taken in the ground-state. For a periodic system the Fourier modes $d_k^\dag$ are given by
\begin{equation}
 d_k^\dag=\frac{1}{\sqrt{L}}\sum_{j=1}^L z^{j(k+1/2)}c_j^\dag\quad,\quad z=e^{2i\pi/L}.
\end{equation}
To simplify Eq.~(\ref{eq:wavedet}) we use (\ref{eq:vand}). We get
\begin{eqnarray}\nonumber
 \psi(x_1,\ldots,x_N)&=&\frac{1}{L^{N/2}}\prod_{j=1}^Nz^{-x_j\frac{N-1}{2}}\prod_{j<k}\left(z^{x_j}-z^{x_k}\right)\\
 &=&\frac{1}{L^{N/2}}\prod_{j>k} 2\sin \frac{\pi (x_k-x_j)}{L},
\end{eqnarray}
up to an unimportant sign. 
The R\'enyi entropy is $S_n=\frac{1}{1-n}\ln Z_n$, where 
\begin{equation}
 Z_n=\frac{1}{N!}\sum_{\{x_i\}}\psi(x_1,\ldots,x_n)^{2n}.
\end{equation}
$Z_n$ is exactly the partition function of a lattice gas with interaction energy
\begin{equation}
 E(x_1,\ldots,x_N)=-\sum_{j<k}\ln \left|z^{x_k}-z^{x_j}\right| +\frac{N}{2}\ln L,
\end{equation}
at inverse temperature $\beta=2n$. The constant energy term can be seen as a positive background charge, and ensures $Z(\beta=2)=Z_1=1$, as expected from the normalization of the ground state. 
This is a discrete analog of the Dyson gas, and has been studied by Gaudin\cite{Gaudin}. Each particle lives on the vertices of $L-$sided regular polygon, as is shown in Fig.~\ref{fig:dysongaudin}(a).

A similar gas also describes the open (Dirichlet) chain. Indeed, in this case the
fermions that diagonalize $H$ are given by
\begin{equation}
 d_k^\dag=\left(\frac{2}{L+1}\right)^{1/2}\sum_{j=1}^L \sin \left(\frac{k\pi j}{L+1}\right) c_j^\dag,
\end{equation}
and we get
\begin{equation}
 \psi(\{x_j\})=\left(\frac{-1}{2L+2}\right)^{N/2}\det_{1\leq j<k\leq N}\left(\omega^{x_j k}-\omega^{-x_j k}\right)
\end{equation}
with $\omega=e^{i\pi/(L+1)}$, 
which can be evaluated using Eq.~(\ref{eq:svand}). The result is
\begin{eqnarray}\nonumber
 \psi(\{x_j\})&=&\left(\frac{-1}{2L+2}\right)^{N/2}\left[\prod_{j=1}^N \omega^{-Nx_j}\left(1-\omega^{2x_j}\right)\right.\\
 &\times&\left.\prod_{k>j}\left(\omega^{x_j}-\omega^{x_k}\right)\left(1-\omega^{x_j}\omega^{x_k}\right)\right].
\end{eqnarray}
$Z_n$ becomes the partition function of a lattice log gas with interaction energy \cite{SMP3}
\begin{eqnarray}\nonumber
 E(\{x_j\})&=&-\sum_{j<k}\ln \left|\omega^{x_j}-\omega^{x_k}\right|-\sum_{j\leq k}\ln \left|\omega^{x_j}-\omega^{-x_k}\right|\\
 &+&\frac{N}{2}\ln (2L+2).
\end{eqnarray}
This is shown in Fig.~\ref{fig:dysongaudin}(b). Each particle now lives on the upper part of the circle, and interacts with the others, its mirror image, and the mirror images of the others.

It turns out a similar representation also exists in the open Neuman case \cite{LazarescuPhD}. Let us denote by $\{y_1,\ldots,y_{L-N}\}=\{1,\ldots,L\}-\{x_1,\ldots,x_N\}$ the positions of the holes. Now we have a new set of $L$ particles: the $N$ original ones to which the $L-N$ mirror images of the holes have been added. The interaction energy is
\begin{eqnarray}\nonumber
 E&=&-\frac{1}{2}\sum_{j< k}^N\ln \left|\omega^{x_j}-\omega^{x_k}\right|-\frac{1}{2}\sum_{j<k}^{L-N}\ln \left|\omega^{-y_j}-\omega^{-y_k}\right|\\
 &-&\frac{1}{2}\sum_{j=1}^{N}\sum_{k=1}^{L-N} \ln \left|\omega^{x_j}-\omega^{-y_k}\right|+\frac{L}{4}\ln (2L+2),
\end{eqnarray}
which is represented in Fig.~\ref{fig:dysongaudin}(c). Since $H$ does not conserve the number of fermions, $N$ is not fixed anymore and the partition function $Z_n=Z(\beta=2n)$ becomes grand-canonical:
\begin{equation}
 Z(\beta)=\sum_{N=0}^{L}\frac{1}{N!}\sum_{x_1,\ldots,x_N} e^{-\beta E(x_1,\ldots,x_N,y_1,\ldots,y_{L-N})}
\end{equation}
\subsection{Summary of the results}
\label{sec:dgresults}
We have seen (see above) that the R\'enyi entropy for three different boundary conditions was given by
\begin{equation}
 S_n=\frac{\ln Z_n}{1-n},
\end{equation}
where $Z_n$ was the partition function of a Dyson-Gaudin gas with certain symmetries (see Fig.~\ref{fig:dysongaudin}(a,b,c)). This correspondence can be exploited to derive exact results in the XX chain. We present them for an XX chain of length $L$, assuming --except for the open Neuman case-- a fixed filling fraction $\rho=N/L$, where $N$ is the number of particles. We also implicitly assume $\rho \leq 1/2$ throughout. The results for $\rho>1/2$ can simply be deduced from the particle-hole symmetry $N\to L-N$.
\subsubsection{Periodic}
For any integer R\'enyi index $n\leq \rho^{-1}$ we have
\begin{equation}\label{eq:gaudin}
 Z_n^{per}(N,L)=\frac{(Nn)!}{N! L^{N(n-1)}n!^N}.
\end{equation}
This result is known \cite{Gaudin} and was exploited in Ref.~[\onlinecite{SFMP}]. Note that all these values belong to the ``replica'' region $n<n_c=\rho^{-2}$. When $n=2$ we also obtained the spin-spin correlation functions. Using the notations of App.~\ref{sec:correlations}, we have for $N\leq L/2$
\begin{eqnarray}\nonumber
 \Braket{\sigma_j^z\sigma_{j+l}^z}_{\!(2)}&=&\frac{2N\cos \frac{2N\pi l}{L}-\cot \frac{\pi l}{L}\sin \frac{2N\pi l}{L}}{L^2\sin \frac{\pi l}{L}}g_N\!\left(\textstyle{\frac{\pi l}{L}}\right)
 \\
 &-&\left[\frac{\sin \frac{2N\pi l}{L}}{L\sin \frac{\pi l}{L}}\right]^2,
 \label{eq:corrxx}
\end{eqnarray}
where 
\begin{equation}
 g_N(\theta)=\sum_{k=1}^{N} \frac{\sin\left[ \left(2k-1\right)\theta\right]}{k-1/2}.
\end{equation}
When $N=L/2$ the result is known \cite{NielsenCiracSierra}, as $\ket{\psi}_{(2)}$ coincides at half-filling with the ground-state of the Haldane-Shastry (HS) chain\cite{Haldane,Shastry}. Hence $Z_2$ is also the norm the (unnormalized) HS ground-state.
In the thermodynamic limit the sum converges to $\pi/2$ for any $l/L$ and $N/L$, so that $\braket{\sigma_j^z\sigma_{j+l}^z}\approx (L\sin\frac{\pi l}{L})^{-1}$. We recover the conformal scaling with a nontrivial exponent $\alpha_2=1$. For a Luttinger liquid the usual one is $\alpha=2K$, and $K=1$ in the XX chain. The exponent $\alpha_2$ we find here is consistent with the image of a modified Luttinger parameter $K^\prime= K/n=K/2=1/2$ near the binding in a replica picture (see Eq.~(\ref{eq:changedstiffness}) and Ref.~[\onlinecite{Kumano}]). 

We were also able to derive a new result for $n=4$ at half-filling:
\begin{equation}\label{eq:periodicspecial}
 Z_4^{per}(L/2,L)=\frac{2^{L/2}}{L^L}\left(\frac{\Gamma\left(\frac{L}{2}+\frac{1}{2}\right)}{\Gamma\left(\frac{1}{2}\right)}\right)^2.
\end{equation}
This is interesting because it corresponds to the critical value of the R\'enyi index. Unfortunately, we did not manage to compute correlation functions at this point. Let us also mention
\begin{equation}\label{eq:sinfty}
S_{\infty}=(\rho \ln 2) L     
\end{equation}
for $\rho\leq 1/2$. All subleading (constant) terms in (\ref{eq:gaudin}, \ref{eq:periodicspecial}, \ref{eq:sinfty}) can be computed, and agree with the predictions of Refs.~[\onlinecite{SFMP,SMP2}].
\subsubsection{Open geometry}
Similar results can be derived in the open geometry. When $n\leq \rho^{-1}$ we have
\begin{equation}\label{eq:openmacdonald}
 Z_n^{open}(N,L)=\left(\frac{n^{n}}{n!(L+1)^{n-1}}\right)^N\times \prod_{k=1}^{n-1}\frac{\Gamma\left(N+\frac{1}{2}+\frac{k}{2n}\right)}{\Gamma\left(\frac{1}{2}+\frac{k}{2n}\right)}
\end{equation}
which, after asymptotic expansion, gives a subleading power-law term $L^{(n-1)/4}$ in $Z_n$, and therefore a $-\frac{1}{4}\ln L$ contribution to the R\'enyi entropy. This is the expected result from CFT \cite{FradkinMoore,Zaletel}. Here also, $Z_2$ coincides with the norm of the Haldane-Shastry chain, with open boundary conditions\cite{BernardPasquierSerban}. 

Just as in the periodic case, we were also able to access $Z_4$ at half-filling. It is given by
\begin{equation}\label{eq:openspecial}
 Z_4^{open}(L/2,L)=\frac{2^{L/2}}{(L+1)^L}\left(\frac{\Gamma \left(\frac{L}{2}+\frac{3}{4}\right)}{\Gamma\left(\frac{3}{4}\right)}\right)^2.
\end{equation}
An asymptotic expansion yields a power-law term $L^{1/2}$, and therefore a contribution $-\frac{1}{6}\ln L$ to the R\'enyi entropy. This is discussed in Sec.~\ref{sec:openchains}. 
\subsubsection{Open Neuman geometry}
The open Neuman chain is slightly more complicated, as the number $N$ of particles is not conserved anymore. We managed to compute $Z_2$, which is given by
\begin{equation}\label{eq:openneuman}
 Z_2=\frac{2^{L/2}}{(L+1)^{L/2+1}}\times \frac{\Gamma\left(\frac{L}{2}+\frac{5}{4}\right)}{\Gamma\left(\frac{5}{4}\right)}.
\end{equation}
An asymptotic expansion gives a power-law term $L^{-1/4}$, and therefore a $+\frac{1}{4}\ln L$ contribution to the second R\'enyi entropy. This is again consistent with CFT \cite{Zaletel}. Also $S_\infty=-\ln p_{\rm max}$ can be computed exactly, with
\begin{eqnarray}\nonumber
 p_{\rm max}&=&\frac{2^{L/2(L/2-1)}}{(L+1)^{L/2}}\left(\cos \frac{\pi}{2L+2}\right)^{-L/2}\\ 
 &\times&\prod_{k=1}^{L/4} \left(\sin \frac{2k\pi}{L+1}\right)^{L-2k}\left(\sin \frac{[2k+1]\pi}{L+1}\right)^{2k}
\end{eqnarray}
(we have assumed $L/2$ even). The asymptotic expansion once again yields a $\frac{1}{4}\ln L$ term, consistent with the Cardy-Peschel formula\cite{CardyPeschel}, with $h_{ND}=1/16$, where $h_{ND}$ is the dimension of the operator that changes the boundary condition from Dirichlet to Neuman\cite{Affleck}.
\subsection{Derivations}
\label{sec:dgdetails}
 Let us now finally explain how the results of the previous subsection can be derived. The techniques are standard in random matrix theory \cite{Forrester}, but there are here a few subtleties \cite{Gaudin} due to the discrete nature of the problem. We focus on the most complicated case, the open geometry, where the results are to our knowledge not known. We have seen that the partition function of the gas is given by
\begin{equation}
 Z_n=\frac{(2L+2)^{-Nn}}{N!}\sum_{\{x\}}\left[ \det_{1\leq j,k\leq N}\left(\omega^{x_j k}-\omega^{-x_j k}\right) \right]^{2n}
\end{equation}
The determinant inside the sum can be evaluated using (\ref{eq:svand}), but we leave it in this form for now. It is easy to check that $Z_1=1$. Indeed for $n=1$ the square can be written as a product of two determinants: explicitly writing the two sum over permutations, and then exchanging the order of the sums yields the result. The case $n=2$ can be obtained for any number of particles. To perform this calculation we rewrite the fourth power of the $\det$ using (\ref{eq:csvand}). We have
\begin{widetext}
\begin{equation}
 Z_2=\frac{(2L+2)^{-2N}}{N!}\sum_{\{x\}}\prod_{j=1}^N \left(w^{2x_j}-1\right)\det \left(\begin{array}{c}
\omega^{x_j k}-\omega^{-x_j k}\\k \omega^{x_j(k-1)}+k \omega^{-x_j(k+1)}
\end{array}
\right),
\end{equation}
where the indices of the $2N\times2N$ determinants run over $j=1,\ldots,N$ and $k=1,\ldots,2N$. Now we expand them using the Laplace formula. After a slight rearrangement we get
\begin{equation}\label{eq:someeq}
 Z_2=\frac{1}{2^N N!}\sum_{\{x\}} \\ \sum_P (-1)^P A(x_1,p_1,p_2) A(x_2,p_3,p_4)\ldots A(x_N,p_{2N-1},p_{2N}),
\end{equation}
where the sum runs over all the permutations $(p_1,p_2,\ldots,p_{2N})$ of $(1,2,\ldots,2N)$, $(-1)^P$ is the signature of the permutation, and
\begin{equation}\label{eq:2by2det}
 A(x,p,q)=\frac{w^{2x}-1}{(2L+2)^2}\det\left(
 \begin{array}{ccc}\omega^{x p}-\omega^{-x p}& \omega^{x q}-\omega^{-x q}\\p [\omega^{x(p-1)}+\omega^{-x(p+1)}]&q [\omega^{x(q-1)}+\omega^{-x(q+1)}]\end{array}
 \right).
\end{equation}
\end{widetext}
Now the sum over all the positions of the particles can be performed, and we obtain
\begin{equation}\label{eq:pfaffian}
 Z_2=\frac{1}{2^NN!}\sum_P(-1)^P M_{p_1 p_2}M_{p_3p_4}\ldots M_{p_{2N-1}p_{2N}},
\end{equation}
where the matrix elements are given by
\begin{equation}\label{eq:truc}
 M_{pq}=\sum_{x=1}^{L}A(x,p,q).
\end{equation}
Eq.~(\ref{eq:pfaffian}) is formally the Pfaffian of the antisymmetric matrix given in (\ref{eq:truc}). It may be calculated by first computing the $2\times 2$ determinant of Eq.~(\ref{eq:2by2det}), and noticing that the matrix elements $M_{pq}$ are non zero only when $p=q\pm 1$. In the end we get, after some further algebra,
\begin{equation}
Z_2(N,L)=\prod_{k=1}^N \frac{4k-1}{2L+2}
 =\left(\frac{2}{L+1}\right)^N\frac{\Gamma\left(N+\frac{3}{4}\right)}{\Gamma\left(\frac{3}{4}\right)}
 \label{eq:open2result}
\end{equation}
when $N\leq L/2$, which is compatible with (\ref{eq:openmacdonald}). When $N>L/2$ we get 
\begin{equation}
Z_2(N,L)=Z_2(L-N,L). 
\end{equation}
 In the periodic case Eqs.~(\ref{eq:pfaffian}) and (\ref{eq:truc}) also hold, with \cite{Gaudin}
\begin{equation}
A(x,p,q)=\frac{q-p}{L^2}z^{x(p+q-2N-1)}.
\end{equation}
The matrix elements are non zero only when $p+q=2N+1$, and using this we reproduce Eq.~(\ref{eq:gaudin}) for $n=2$.

When $n$ is greater than two we use a different method, and rely on the result of Macdonald explained in Sec.~\ref{sec:dysonmacdonald}. In an open system the partition function can be written, using (\ref{eq:svand}), as
\begin{widetext}
\begin{equation}\label{eq:part}
 Z_n=\frac{(2L+2)^{-Nn}}{N!}\sum_{\{x_j\}}
 \left[
 \prod_{j=1}^N\left(1-\omega^{2x_j}\right)\left(1-\omega^{-2x_j}\right)
\prod_{1\leq j<k\leq N}\left(1-\omega^{x_j-x_k}\right)\left(1-\omega^{x_k-x_j}\right)\left(1-\omega^{x_j+x_k}\right)\left(1-\omega^{-x_j-x_k}\right)
 \right]^n
\end{equation}
\end{widetext}
Now if we expand the $n$-th power of the product in (\ref{eq:part}), we get a sum of terms of the form $\omega^{\sum_i m_i x_i}$ for some integers $m_i$. The crucial point is that each $m_i$ has its module bounded by $|m_i|\leq 2nN$. Since $\sum_{x=1}^{L} \omega^{mx}=0$ unless $m$ is zero or a multiple of $2L+2$, only the constant term of the expanded product has a non vanishing contribution in the sum, \emph{provided} $n<(L+1)/N$. This constant term is exactly given by the Macdonald result for the $C_N$ root systems, as is explained in section \ref{sec:dysonmacdonald}. Plugging (\ref{eq:di}) and (\ref{eq:macdonald}) in (\ref{eq:part}) gives our result Eq.~(\ref{eq:openmacdonald}). The periodic case follows from a similar argument \cite{Gaudin} with the constant term identity corresponding to the $A_{N-1}$ root system, also known as the Dyson conjecture. We recover Eq.~(\ref{eq:gaudin}), provided $n<L/(N-1)$. The restriction on the allowed values of $n$ is an important difference with the continuous Dyson gas. At filling half or less, $n=2$ is in the allowed values of $n$, and we also reproduce Eq.~(\ref{eq:open2result}) which we derived first using the determinant approach. The latter method has an interesting advantage, in that it can easily be extended to compute correlation functions at $n=2$. Indeed, let us for example look at the two point correlation function of two particles at position $x_1$ and $x_2$:
\begin{equation}
 C_{2}(x_1,x_2)=\frac{\sum_{\{x\}}^\prime \psi(x_1,\ldots,x_N)^4}{\sum_{\{x\}}\psi(x_1,\ldots,x_N)^4}
\end{equation}
where in $\sum^\prime_{\{x\}}$ we sum over all the position $x_3,x_4,\ldots,x_N$ of the remaining particles. The calculation is easiest in the periodic case. Using the approach leading to (\ref{eq:someeq}), only few permutations give a nonzero contribution. For that to happen $(p_1,p_2,p_3,p_4)$ has to be a permutation of $(k,2N+1-k,k^\prime,2N+1-k^\prime)$ for some choice of $1\leq k,k^\prime\leq N$. We get
\begin{equation}\label{eq:twopoints1}
 C_{2}(x_1,x_2)=\frac{L^2}{4}\sum_{1\leq k<k^\prime\leq N}\frac{\Upsilon_{kk^\prime}(x_1,x_2)}{(2N+1-2k)(2N+1-2k^\prime)},
\end{equation}
where
\begin{equation}\label{eq:twopoints2}
 \Upsilon_{kk^\prime}(x_1,x_2)=\sum_{Q\in S_4} (-1)^Q A(x_1,q_1,q_2)A(x_2,q_3,q_4).
\end{equation}
Here the sum runs over all $4!=24$ permutations of $(k,2N+1-k,k^\prime,2N+1-k^\prime)$, and can be computed explicitly. In the end we recover (\ref{eq:corrxx}), using $\braket{\sigma_{x_1}^z\sigma_{x_2}^z}_{\!(2)}=4C_{2}(x_1,x_2)-4N^2/L^2$, and further simplifications. 

The determinant method also allows to access $Z_4$, using the confluent versions of Vandermonde (\ref{eq:cvand},\ref{eq:csvand}) and expanding the product of the two determinants similar to the calculation of $Z_1$. The computations are somewhat cumbersome, but we can nevertheless recover the results obtained through the Macdonald conjecture (when $\rho \leq 1/4$). Outside of this region we could not obtain a simple closed-form formula, \emph{except} in the special case of half-filling. This is interesting, as the  Macdonald conjecture does not give the result. The method works for both periodic and open (Dirichlet) systems. The results are given by Eq.~(\ref{eq:periodicspecial}) and Eq.~(\ref{eq:openspecial}), and correspond exactly to the transition point $n=n_c$ discussed in the text. 

In the open Neuman case this method is the only one available, as the grand-canonical nature of the partition function necessarily breaks the condition $n\leq \rho^{-1}$, and the constant term results cannot be used anymore. The result for $Z_2$ is given by (\ref{eq:openneuman}). 

Let us finally mention that $S_\infty$ can be obtained by noticing that the most likely configurations are attained for particles as far apart from each other as possible. For example at half-filling this corresponds to the configuration $(1,3,5,\ldots ,L-1)$ and $(2,4,6,\ldots ,L)$ for the positions of the particles. In spin language, these are the homogeneous states $\ket{\uparrow\downarrow\ldots\uparrow\downarrow}$ and $\ket{\downarrow\uparrow\ldots\downarrow\uparrow}$ discussed in the main text.

 \bibliography{SMI4}

\begin{thebibliography}{64}%
\makeatletter
\providecommand \@ifxundefined [1]{%
 \@ifx{#1\undefined}
}%
\providecommand \@ifnum [1]{%
 \ifnum #1\expandafter \@firstoftwo
 \else \expandafter \@secondoftwo
 \fi
}%
\providecommand \@ifx [1]{%
 \ifx #1\expandafter \@firstoftwo
 \else \expandafter \@secondoftwo
 \fi
}%
\providecommand \natexlab [1]{#1}%
\providecommand \enquote  [1]{``#1''}%
\providecommand \bibnamefont  [1]{#1}%
\providecommand \bibfnamefont [1]{#1}%
\providecommand \citenamefont [1]{#1}%
\providecommand \href@noop [0]{\@secondoftwo}%
\providecommand \href [0]{\begingroup \@sanitize@url \@href}%
\providecommand \@href[1]{\@@startlink{#1}\@@href}%
\providecommand \@@href[1]{\endgroup#1\@@endlink}%
\providecommand \@sanitize@url [0]{\catcode `\\12\catcode `\$12\catcode
  `\&12\catcode `\#12\catcode `\^12\catcode `\_12\catcode `\%12\relax}%
\providecommand \@@startlink[1]{}%
\providecommand \@@endlink[0]{}%
\providecommand \url  [0]{\begingroup\@sanitize@url \@url }%
\providecommand \@url [1]{\endgroup\@href {#1}{\urlprefix }}%
\providecommand \urlprefix  [0]{URL }%
\providecommand \Eprint [0]{\href }%
\providecommand \doibase [0]{http://dx.doi.org/}%
\providecommand \selectlanguage [0]{\@gobble}%
\providecommand \bibinfo  [0]{\@secondoftwo}%
\providecommand \bibfield  [0]{\@secondoftwo}%
\providecommand \translation [1]{[#1]}%
\providecommand \BibitemOpen [0]{}%
\providecommand \bibitemStop [0]{}%
\providecommand \bibitemNoStop [0]{.\EOS\space}%
\providecommand \EOS [0]{\spacefactor3000\relax}%
\providecommand \BibitemShut  [1]{\csname bibitem#1\endcsname}%
\let\auto@bib@innerbib\@empty
\bibitem [{\citenamefont {Alcaraz}\ and\ \citenamefont
  {Rajabpour}(2013)}]{AlcarazRajabpour}%
  \BibitemOpen
  \bibfield  {author} {\bibinfo {author} {\bibfnamefont {F.~C.}\ \bibnamefont
  {Alcaraz}}\ and\ \bibinfo {author} {\bibfnamefont {M.~A.}\ \bibnamefont
  {Rajabpour}},\ }\href {\doibase 10.1103/PhysRevLett.111.017201} {\bibfield
  {journal} {\bibinfo  {journal} {Phys. Rev. Lett.}\ }\textbf {\bibinfo
  {volume} {111}},\ \bibinfo {pages} {017201} (\bibinfo {year} {2013})},\
  \Eprint {http://arxiv.org/abs/1305.1239} {arXiv:1305.1239} \BibitemShut
  {NoStop}%
\bibitem [{\citenamefont {Holzhey}\ \emph {et~al.}(1994)\citenamefont
  {Holzhey}, \citenamefont {Larsen},\ and\ \citenamefont {Wilczek}}]{EE1d1}%
  \BibitemOpen
  \bibfield  {author} {\bibinfo {author} {\bibfnamefont {C.}~\bibnamefont
  {Holzhey}}, \bibinfo {author} {\bibfnamefont {F.}~\bibnamefont {Larsen}}, \
  and\ \bibinfo {author} {\bibfnamefont {F.}~\bibnamefont {Wilczek}},\ }\href
  {\doibase 10.1016/0550-3213(94)90402-2} {\bibfield  {journal} {\bibinfo
  {journal} {Nucl.Phys. B}\ }\textbf {\bibinfo {volume} {424}},\ \bibinfo
  {pages} {443} (\bibinfo {year} {1994})},\ \Eprint
  {http://arxiv.org/abs/hep-th/9403108} {arXiv:hep-th/9403108} \BibitemShut
  {NoStop}%
\bibitem [{\citenamefont {Vidal}\ \emph {et~al.}(2003)\citenamefont {Vidal},
  \citenamefont {Latorre}, \citenamefont {Rico},\ and\ \citenamefont
  {Kitaev}}]{EE1d2}%
  \BibitemOpen
  \bibfield  {author} {\bibinfo {author} {\bibfnamefont {G.}~\bibnamefont
  {Vidal}}, \bibinfo {author} {\bibfnamefont {J.~I.}\ \bibnamefont {Latorre}},
  \bibinfo {author} {\bibfnamefont {E.}~\bibnamefont {Rico}}, \ and\ \bibinfo
  {author} {\bibfnamefont {A.}~\bibnamefont {Kitaev}},\ }\href {\doibase
  10.1103/PhysRevLett.90.227902} {\bibfield  {journal} {\bibinfo  {journal}
  {Phys.Rev.Lett.}\ }\textbf {\bibinfo {volume} {90}},\ \bibinfo {pages}
  {227902} (\bibinfo {year} {2003})},\ \Eprint
  {http://arxiv.org/abs/quant-ph/0211074} {arXiv:quant-ph/0211074} \BibitemShut
  {NoStop}%
\bibitem [{\citenamefont {Calabrese}\ and\ \citenamefont
  {Cardy}(2004)}]{EE1d3}%
  \BibitemOpen
  \bibfield  {author} {\bibinfo {author} {\bibfnamefont {P.}~\bibnamefont
  {Calabrese}}\ and\ \bibinfo {author} {\bibfnamefont {J.}~\bibnamefont
  {Cardy}},\ }\href {http://dx.doi.org/10.1088/1742-5468/2004/06/P06002}
  {\bibfield  {journal} {\bibinfo  {journal} {J.Stat.Mech. P06002}\ } (\bibinfo
  {year} {2004})},\ \Eprint {http://arxiv.org/abs/hep-th/0405152}
  {arXiv:hep-th/0405152} \BibitemShut {NoStop}%
\bibitem [{\citenamefont {Belavin}\ \emph {et~al.}(1984)\citenamefont
  {Belavin}, \citenamefont {Polyakov},\ and\ \citenamefont
  {Zamolodchikov}}]{BPZ}%
  \BibitemOpen
  \bibfield  {author} {\bibinfo {author} {\bibfnamefont {A.~A.}\ \bibnamefont
  {Belavin}}, \bibinfo {author} {\bibfnamefont {A.~M.}\ \bibnamefont
  {Polyakov}}, \ and\ \bibinfo {author} {\bibfnamefont {A.~B.}\ \bibnamefont
  {Zamolodchikov}},\ }\href {http://dx.doi.org/10.1016/0550-3213(84)90052-X}
  {\bibfield  {journal} {\bibinfo  {journal} {Nucl. Phys. B}\ }\textbf
  {\bibinfo {volume} {241}},\ \bibinfo {pages} {333} (\bibinfo {year}
  {1984})}\BibitemShut {NoStop}%
\bibitem [{\citenamefont {{Di Franceso}}\ \emph {et~al.}(1997)\citenamefont
  {{Di Franceso}}, \citenamefont {Mathieu},\ and\ \citenamefont
  {S{\'e}n{\'e}chal}}]{Yellowpages}%
  \BibitemOpen
  \bibfield  {author} {\bibinfo {author} {\bibfnamefont {P.}~\bibnamefont {{Di
  Franceso}}}, \bibinfo {author} {\bibfnamefont {P.}~\bibnamefont {Mathieu}}, \
  and\ \bibinfo {author} {\bibfnamefont {D.}~\bibnamefont {S{\'e}n{\'e}chal}},\
  }\href@noop {} {\emph {\bibinfo {title} {{Conformal Field Theory}}}}\
  (\bibinfo  {publisher} {Springer-verlag},\ \bibinfo {year}
  {1997})\BibitemShut {NoStop}%
\bibitem [{\citenamefont {Calabrese}\ \emph {et~al.}(2012)\citenamefont
  {Calabrese}, \citenamefont {Cardy},\ and\ \citenamefont
  {Tonni}}]{negativity}%
  \BibitemOpen
  \bibfield  {author} {\bibinfo {author} {\bibfnamefont {P.}~\bibnamefont
  {Calabrese}}, \bibinfo {author} {\bibfnamefont {J.}~\bibnamefont {Cardy}}, \
  and\ \bibinfo {author} {\bibfnamefont {E.}~\bibnamefont {Tonni}},\ }\href
  {\doibase 10.1103/PhysRevLett.109.130502} {\bibfield  {journal} {\bibinfo
  {journal} {Phys. Rev. Lett.}\ }\textbf {\bibinfo {volume} {109}},\ \bibinfo
  {pages} {130502} (\bibinfo {year} {2012})},\ \Eprint
  {http://arxiv.org/abs/1206.3092} {arXiv:1206.3092} \BibitemShut {NoStop}%
\bibitem [{\citenamefont {Dubail}\ and\ \citenamefont
  {St{\'e}phan}(2011)}]{LBF}%
  \BibitemOpen
  \bibfield  {author} {\bibinfo {author} {\bibfnamefont {J.}~\bibnamefont
  {Dubail}}\ and\ \bibinfo {author} {\bibfnamefont {J.-M.}\ \bibnamefont
  {St{\'e}phan}},\ }\href {http://dx.doi.org/10.1088/1742-5468/2011/03/L03002}
  {\bibfield  {journal} {\bibinfo  {journal} {J. Stat. Mech. L03002}\ }
  (\bibinfo {year} {2011})},\ \Eprint {http://arxiv.org/abs/1010.3716}
  {arXiv:1010.3716} \BibitemShut {NoStop}%
\bibitem [{\citenamefont {Evers}\ and\ \citenamefont
  {Mirlin}(2008)}]{ErversMirlin}%
  \BibitemOpen
  \bibfield  {author} {\bibinfo {author} {\bibfnamefont {F.}~\bibnamefont
  {Evers}}\ and\ \bibinfo {author} {\bibfnamefont {A.~D.}\ \bibnamefont
  {Mirlin}},\ }\href {\doibase 10.1103/RevModPhys.80.1355} {\bibfield
  {journal} {\bibinfo  {journal} {Rev. Mod. Phys. 1355}\ }\textbf {\bibinfo
  {volume} {80}},\ \bibinfo {pages} {1355} (\bibinfo {year} {2008})},\ \Eprint
  {http://arxiv.org/abs/0707.4378} {arXiv:0707.4378} \BibitemShut {NoStop}%
\bibitem [{\citenamefont {Evers}\ and\ \citenamefont
  {Mirlin}(2000)}]{ErversMirlinPRL}%
  \BibitemOpen
  \bibfield  {author} {\bibinfo {author} {\bibfnamefont {F.}~\bibnamefont
  {Evers}}\ and\ \bibinfo {author} {\bibfnamefont {A.~D.}\ \bibnamefont
  {Mirlin}},\ }\href {\doibase 10.1103/PhysRevLett.84.3690} {\bibfield
  {journal} {\bibinfo  {journal} {Phys. Rev. Lett.}\ }\textbf {\bibinfo
  {volume} {84}},\ \bibinfo {pages} {3690} (\bibinfo {year} {2000})},\ \Eprint
  {http://arxiv.org/abs/cond-mat/0001086} {arXiv:cond-mat/0001086} \BibitemShut
  {NoStop}%
\bibitem [{\citenamefont {Atas}\ and\ \citenamefont
  {Bogomolny}(2012)}]{AtasBogomolny}%
  \BibitemOpen
  \bibfield  {author} {\bibinfo {author} {\bibfnamefont {Y.~Y.}\ \bibnamefont
  {Atas}}\ and\ \bibinfo {author} {\bibfnamefont {E.}~\bibnamefont
  {Bogomolny}},\ }\href {http://dx.doi.org/10.1103/PhysRevE.86.021104}
  {\bibfield  {journal} {\bibinfo  {journal} {Phys. Rev. E}\ }\textbf {\bibinfo
  {volume} {86}},\ \bibinfo {pages} {021104} (\bibinfo {year} {2012})},\
  \Eprint {http://arxiv.org/abs/1205.4541} {arXiv:1205.4541} \BibitemShut
  {NoStop}%
\bibitem [{\citenamefont {Sinai}(1968)}]{Sinai}%
  \BibitemOpen
  \bibfield  {author} {\bibinfo {author} {\bibfnamefont {Y.~G.}\ \bibnamefont
  {Sinai}},\ }\href {\doibase 10.1007/BF01075361} {\bibfield  {journal}
  {\bibinfo  {journal} {Funct. Anal. Appl.}\ }\textbf {\bibinfo {volume}
  {24}},\ \bibinfo {pages} {61} (\bibinfo {year} {1968})}\BibitemShut {NoStop}%
\bibitem [{\citenamefont {Israilev}(1990)}]{Izrailev}%
  \BibitemOpen
  \bibfield  {author} {\bibinfo {author} {\bibfnamefont {F.~M.}\ \bibnamefont
  {Israilev}},\ }\href {http://dx.doi.org/10.1016/0370-1573(90)90067-C}
  {\bibfield  {journal} {\bibinfo  {journal} {Physics Reports}\ }\textbf
  {\bibinfo {volume} {196}},\ \bibinfo {pages} {299} (\bibinfo {year}
  {1990})}\BibitemShut {NoStop}%
\bibitem [{\citenamefont {Santos}\ and\ \citenamefont
  {Rigol}(2010)}]{SantosRigol}%
  \BibitemOpen
  \bibfield  {author} {\bibinfo {author} {\bibfnamefont {L.~F.}\ \bibnamefont
  {Santos}}\ and\ \bibinfo {author} {\bibfnamefont {M.}~\bibnamefont {Rigol}},\
  }\href {\doibase 10.1103/PhysRevE.81.036206} {\bibfield  {journal} {\bibinfo
  {journal} {Phys. Rev. E}\ }\textbf {\bibinfo {volume} {81}},\ \bibinfo
  {pages} {036206} (\bibinfo {year} {2010})},\ \Eprint
  {http://arxiv.org/abs/0910.2985} {arXiv:0910.2985} \BibitemShut {NoStop}%
\bibitem [{\citenamefont {St{\'e}phan}\ \emph {et~al.}(2009)\citenamefont
  {St{\'e}phan}, \citenamefont {Furukawa}, \citenamefont {Misguich},\ and\
  \citenamefont {Pasquier}}]{SFMP}%
  \BibitemOpen
  \bibfield  {author} {\bibinfo {author} {\bibfnamefont {J.-M.}\ \bibnamefont
  {St{\'e}phan}}, \bibinfo {author} {\bibfnamefont {S.}~\bibnamefont
  {Furukawa}}, \bibinfo {author} {\bibfnamefont {G.}~\bibnamefont {Misguich}},
  \ and\ \bibinfo {author} {\bibfnamefont {V.}~\bibnamefont {Pasquier}},\
  }\href {\doibase 10.1103/PhysRevB.80.184421} {\bibfield  {journal} {\bibinfo
  {journal} {Phys. Rev. B}\ }\textbf {\bibinfo {volume} {80}},\ \bibinfo
  {pages} {184421} (\bibinfo {year} {2009})},\ \Eprint
  {http://arxiv.org/abs/0906.1153} {arXiv:0906.1153} \BibitemShut {NoStop}%
\bibitem [{\citenamefont {Fradkin}\ and\ \citenamefont
  {Moore}(2006)}]{FradkinMoore}%
  \BibitemOpen
  \bibfield  {author} {\bibinfo {author} {\bibfnamefont {E.}~\bibnamefont
  {Fradkin}}\ and\ \bibinfo {author} {\bibfnamefont {J.~E.}\ \bibnamefont
  {Moore}},\ }\href {http://dx.doi.org/10.1103/PhysRevLett.97.050404}
  {\bibfield  {journal} {\bibinfo  {journal} {Phys.Rev.Lett.}\ }\textbf
  {\bibinfo {volume} {97}},\ \bibinfo {pages} {050404} (\bibinfo {year}
  {2006})},\ \Eprint {http://arxiv.org/abs/cond-mat/0605683}
  {arXiv:cond-mat/0605683} \BibitemShut {NoStop}%
\bibitem [{\citenamefont {Hsu}\ \emph {et~al.}(2009)\citenamefont {Hsu},
  \citenamefont {Mulligan}, \citenamefont {Fradkin},\ and\ \citenamefont
  {Kim}}]{Hsuetal}%
  \BibitemOpen
  \bibfield  {author} {\bibinfo {author} {\bibfnamefont {B.}~\bibnamefont
  {Hsu}}, \bibinfo {author} {\bibfnamefont {M.}~\bibnamefont {Mulligan}},
  \bibinfo {author} {\bibfnamefont {E.}~\bibnamefont {Fradkin}}, \ and\
  \bibinfo {author} {\bibfnamefont {E.-A.}\ \bibnamefont {Kim}},\ }\href
  {\doibase 10.1103/PhysRevB.79.115421} {\bibfield  {journal} {\bibinfo
  {journal} {Phys. Rev. B}\ }\textbf {\bibinfo {volume} {79}},\ \bibinfo
  {pages} {115421} (\bibinfo {year} {2009})},\ \Eprint
  {http://arxiv.org/abs/0812.0203} {arXiv:0812.0203} \BibitemShut {NoStop}%
\bibitem [{\citenamefont {St{\'e}phan}\ \emph {et~al.}(2010)\citenamefont
  {St{\'e}phan}, \citenamefont {Misguich},\ and\ \citenamefont
  {Pasquier}}]{SMP2}%
  \BibitemOpen
  \bibfield  {author} {\bibinfo {author} {\bibfnamefont {J.-M.}\ \bibnamefont
  {St{\'e}phan}}, \bibinfo {author} {\bibfnamefont {G.}~\bibnamefont
  {Misguich}}, \ and\ \bibinfo {author} {\bibfnamefont {V.}~\bibnamefont
  {Pasquier}},\ }\href {\doibase 10.1103/PhysRevB.82.125455} {\bibfield
  {journal} {\bibinfo  {journal} {Phys. Rev. B}\ }\textbf {\bibinfo {volume}
  {82}},\ \bibinfo {pages} {125455} (\bibinfo {year} {2010})},\ \Eprint
  {http://arxiv.org/abs/1006.1605} {arXiv:1006.1605} \BibitemShut {NoStop}%
\bibitem [{\citenamefont {Oshikawa}(2010)}]{Oshikawa}%
  \BibitemOpen
  \bibfield  {author} {\bibinfo {author} {\bibfnamefont {M.}~\bibnamefont
  {Oshikawa}},\ }\href@noop {} {\bibfield  {journal} {\bibinfo  {journal}
  {preprint}\ } (\bibinfo {year} {2010})},\ \Eprint
  {http://arxiv.org/abs/1007.3739} {arXiv:1007.3739} \BibitemShut {NoStop}%
\bibitem [{\citenamefont {Zaletel}\ \emph {et~al.}(2011)\citenamefont
  {Zaletel}, \citenamefont {Bardarson},\ and\ \citenamefont {Moore}}]{Zaletel}%
  \BibitemOpen
  \bibfield  {author} {\bibinfo {author} {\bibfnamefont {M.~P.}\ \bibnamefont
  {Zaletel}}, \bibinfo {author} {\bibfnamefont {J.~H.}\ \bibnamefont
  {Bardarson}}, \ and\ \bibinfo {author} {\bibfnamefont {J.~E.}\ \bibnamefont
  {Moore}},\ }\href {http://dx.doi.org/10.1103/PhysRevLett.107.020402}
  {\bibfield  {journal} {\bibinfo  {journal} {Phys. Rev. Lett.}\ }\textbf
  {\bibinfo {volume} {107}},\ \bibinfo {pages} {020402} (\bibinfo {year}
  {2011})},\ \Eprint {http://arxiv.org/abs/1103.5452} {arXiv:1103.5452}
  \BibitemShut {NoStop}%
\bibitem [{\citenamefont {St{\'e}phan}\ \emph {et~al.}(2011)\citenamefont
  {St{\'e}phan}, \citenamefont {Misguich},\ and\ \citenamefont
  {Pasquier}}]{SMP3}%
  \BibitemOpen
  \bibfield  {author} {\bibinfo {author} {\bibfnamefont {J.-M.}\ \bibnamefont
  {St{\'e}phan}}, \bibinfo {author} {\bibfnamefont {G.}~\bibnamefont
  {Misguich}}, \ and\ \bibinfo {author} {\bibfnamefont {V.}~\bibnamefont
  {Pasquier}},\ }\href {\doibase 10.1103/PhysRevB.84.195128} {\bibfield
  {journal} {\bibinfo  {journal} {Phys. Rev. B}\ }\textbf {\bibinfo {volume}
  {84}},\ \bibinfo {pages} {195128} (\bibinfo {year} {2011})},\ \Eprint
  {http://arxiv.org/abs/1104.2544} {arXiv:1104.2544} \BibitemShut {NoStop}%
\bibitem [{\citenamefont {Ju}\ \emph {et~al.}(2012)\citenamefont {Ju},
  \citenamefont {Kallin}, \citenamefont {Fendley}, \citenamefont {Hastings},\
  and\ \citenamefont {Melko}}]{RVB1}%
  \BibitemOpen
  \bibfield  {author} {\bibinfo {author} {\bibfnamefont {H.}~\bibnamefont
  {Ju}}, \bibinfo {author} {\bibfnamefont {A.~B.}\ \bibnamefont {Kallin}},
  \bibinfo {author} {\bibfnamefont {P.}~\bibnamefont {Fendley}}, \bibinfo
  {author} {\bibfnamefont {M.~B.}\ \bibnamefont {Hastings}}, \ and\ \bibinfo
  {author} {\bibfnamefont {R.~G.}\ \bibnamefont {Melko}},\ }\href {\doibase
  10.1103/PhysRevB.85.165121} {\bibfield  {journal} {\bibinfo  {journal} {Phys.
  Rev. B 85}\ }\textbf {\bibinfo {volume} {85}},\ \bibinfo {pages} {165121}
  (\bibinfo {year} {2012})},\ \Eprint {http://arxiv.org/abs/1112.4474}
  {arXiv:1112.4474} \BibitemShut {NoStop}%
\bibitem [{\citenamefont {St{\'e}phan}\ \emph {et~al.}(2013)\citenamefont
  {St{\'e}phan}, \citenamefont {Ju}, \citenamefont {Fendley},\ and\
  \citenamefont {Melko}}]{RVB2}%
  \BibitemOpen
  \bibfield  {author} {\bibinfo {author} {\bibfnamefont {J.-M.}\ \bibnamefont
  {St{\'e}phan}}, \bibinfo {author} {\bibfnamefont {H.}~\bibnamefont {Ju}},
  \bibinfo {author} {\bibfnamefont {P.}~\bibnamefont {Fendley}}, \ and\
  \bibinfo {author} {\bibfnamefont {R.~G.}\ \bibnamefont {Melko}},\ }\href
  {\doibase 10.1088/1367-2630/15/1/015004} {\bibfield  {journal} {\bibinfo
  {journal} {New. J. Phys.}\ }\textbf {\bibinfo {volume} {15}},\ \bibinfo
  {pages} {015004} (\bibinfo {year} {2013})},\ \Eprint
  {http://arxiv.org/abs/1207.3820} {arXiv:1207.3820} \BibitemShut {NoStop}%
\bibitem [{\citenamefont {St{\'e}phan}\ \emph {et~al.}(2014)\citenamefont
  {St{\'e}phan}, \citenamefont {Inglis}, \citenamefont {Fendley},\ and\
  \citenamefont {Melko}}]{GMI}%
  \BibitemOpen
  \bibfield  {author} {\bibinfo {author} {\bibfnamefont {J.-M.}\ \bibnamefont
  {St{\'e}phan}}, \bibinfo {author} {\bibfnamefont {S.}~\bibnamefont {Inglis}},
  \bibinfo {author} {\bibfnamefont {P.}~\bibnamefont {Fendley}}, \ and\
  \bibinfo {author} {\bibfnamefont {R.~G.}\ \bibnamefont {Melko}},\ }\href
  {http://dx.doi.org/10.1103/PhysRevLett.112.127204} {\bibfield  {journal}
  {\bibinfo  {journal} {Phys. Rev. Lett.}\ }\textbf {\bibinfo {volume} {112}},\
  \bibinfo {pages} {127204} (\bibinfo {year} {2014})},\ \Eprint
  {http://arxiv.org/abs/1312.3954} {arXiv:1312.3954} \BibitemShut {NoStop}%
\bibitem [{Note1()}]{Note1}%
  \BibitemOpen
  \bibinfo {note} {The volume law may be violated if the locality requirement
  is dropped. For example the Shannon entropy evaluated in the basis of the
  eigenstates of the Hamiltonian --which is usually highly nonlocal-- is $0$ if
  the ground-state is nondegenerate.}\BibitemShut {Stop}%
\bibitem [{\citenamefont {Luitz}\ \emph
  {et~al.}(2014{\natexlab{a}})\citenamefont {Luitz}, \citenamefont
  {Laflorencie},\ and\ \citenamefont {Alet}}]{AletShannon3}%
  \BibitemOpen
  \bibfield  {author} {\bibinfo {author} {\bibfnamefont {D.~J.}\ \bibnamefont
  {Luitz}}, \bibinfo {author} {\bibfnamefont {N.}~\bibnamefont {Laflorencie}},
  \ and\ \bibinfo {author} {\bibfnamefont {F.}~\bibnamefont {Alet}},\ }\href
  {http://arxiv.org/abs/1404.3717v1; http://arxiv.org/pdf/1404.3717v1} {\
  (\bibinfo {year} {2014}{\natexlab{a}})},\ \Eprint
  {http://arxiv.org/abs/1404.3717} {arXiv:1404.3717} \BibitemShut {NoStop}%
\bibitem [{\citenamefont {Luitz}\ \emph
  {et~al.}(2014{\natexlab{b}})\citenamefont {Luitz}, \citenamefont {Alet},\
  and\ \citenamefont {Laflorencie}}]{AletShannon}%
  \BibitemOpen
  \bibfield  {author} {\bibinfo {author} {\bibfnamefont {D.~J.}\ \bibnamefont
  {Luitz}}, \bibinfo {author} {\bibfnamefont {F.}~\bibnamefont {Alet}}, \ and\
  \bibinfo {author} {\bibfnamefont {N.}~\bibnamefont {Laflorencie}},\ }\href
  {http://dx.doi.org/10.1103/PhysRevLett.112.057203} {\bibfield  {journal}
  {\bibinfo  {journal} {Phys. Rev. Lett.}\ }\textbf {\bibinfo {volume} {112}},\
  \bibinfo {pages} {057203} (\bibinfo {year} {2014}{\natexlab{b}})},\ \Eprint
  {http://arxiv.org/abs/1308.1916} {arXiv:1308.1916} \BibitemShut {NoStop}%
\bibitem [{\citenamefont {Luitz}\ \emph
  {et~al.}(2014{\natexlab{c}})\citenamefont {Luitz}, \citenamefont {Alet},\
  and\ \citenamefont {Laflorencie}}]{AletShannon2}%
  \BibitemOpen
  \bibfield  {author} {\bibinfo {author} {\bibfnamefont {D.~J.}\ \bibnamefont
  {Luitz}}, \bibinfo {author} {\bibfnamefont {F.}~\bibnamefont {Alet}}, \ and\
  \bibinfo {author} {\bibfnamefont {N.}~\bibnamefont {Laflorencie}},\ }\href
  {http://dx.doi.org/10.1103/PhysRevB.89.165106} {\bibfield  {journal}
  {\bibinfo  {journal} {Phys. Rev. B}\ }\textbf {\bibinfo {volume} {89}},\
  \bibinfo {pages} {165106} (\bibinfo {year} {2014}{\natexlab{c}})},\ \Eprint
  {http://arxiv.org/abs/1402.4813} {arXiv:1402.4813} \BibitemShut {NoStop}%
\bibitem [{\citenamefont {Um}\ \emph {et~al.}(2012)\citenamefont {Um},
  \citenamefont {Park},\ and\ \citenamefont {Hinrichsen}}]{Umetal}%
  \BibitemOpen
  \bibfield  {author} {\bibinfo {author} {\bibfnamefont {J.}~\bibnamefont
  {Um}}, \bibinfo {author} {\bibfnamefont {H.}~\bibnamefont {Park}}, \ and\
  \bibinfo {author} {\bibfnamefont {H.}~\bibnamefont {Hinrichsen}},\ }\href
  {http://dx.doi.org/10.1088/1742-5468/2012/10/P10026} {\bibfield  {journal}
  {\bibinfo  {journal} {J. Stat. Mech. P10026}\ } (\bibinfo {year} {2012})},\
  \Eprint {http://arxiv.org/abs/1208.4962} {arXiv:1208.4962} \BibitemShut
  {NoStop}%
\bibitem [{\citenamefont {Lau}\ and\ \citenamefont
  {Grassberger}(2012)}]{LauGrassberger}%
  \BibitemOpen
  \bibfield  {author} {\bibinfo {author} {\bibfnamefont {H.~W.}\ \bibnamefont
  {Lau}}\ and\ \bibinfo {author} {\bibfnamefont {P.}~\bibnamefont
  {Grassberger}},\ }\href {\doibase 10.1103/PhysRevE.87.022128} {\bibfield
  {journal} {\bibinfo  {journal} {Phys. Rev. E}\ }\textbf {\bibinfo {volume}
  {87}},\ \bibinfo {pages} {022128} (\bibinfo {year} {2012})},\ \Eprint
  {http://arxiv.org/abs/1210.5707} {arXiv:1210.5707} \BibitemShut {NoStop}%
\bibitem [{\citenamefont {St{\'e}phan}(2011)}]{PhDStephan}%
  \BibitemOpen
  \bibfield  {author} {\bibinfo {author} {\bibfnamefont {J.-M.}\ \bibnamefont
  {St{\'e}phan}},\ }\emph {\bibinfo {title} {{Intrication dans des syst{\`e}mes
  quantiques {\`a} basse dimension}}},\ \href
  {http://ipht.cea.fr/Docspht//articles/t11/231/public/these-stephan.pdf}
  {Ph.D. thesis} (\bibinfo {year} {p87-89, in French, 2011})\BibitemShut
  {NoStop}%
\bibitem [{\citenamefont {Cardy}(1989)}]{Cardybcc}%
  \BibitemOpen
  \bibfield  {author} {\bibinfo {author} {\bibfnamefont {J.~L.}\ \bibnamefont
  {Cardy}},\ }\href {http://dx.doi.org/10.1016/0550-3213(89)90521-X} {\bibfield
   {journal} {\bibinfo  {journal} {Nucl. Phys. B}\ }\textbf {\bibinfo {volume}
  {324}},\ \bibinfo {pages} {581} (\bibinfo {year} {1989})}\BibitemShut
  {NoStop}%
\bibitem [{\citenamefont {St{\'e}phan}(2014)}]{emptiness}%
  \BibitemOpen
  \bibfield  {author} {\bibinfo {author} {\bibfnamefont {J.-M.}\ \bibnamefont
  {St{\'e}phan}},\ }\href {http://dx.doi.org/10.1088/1742-5468/2014/05/P05010}
  {\bibfield  {journal} {\bibinfo  {journal} {J. Stat. Mech P05010}\ }
  (\bibinfo {year} {2014})},\ \Eprint {http://arxiv.org/abs/1303.5499}
  {arXiv:1303.5499} \BibitemShut {NoStop}%
\bibitem [{\citenamefont {Cardy}\ and\ \citenamefont
  {Peschel}(1988)}]{CardyPeschel}%
  \BibitemOpen
  \bibfield  {author} {\bibinfo {author} {\bibfnamefont {J.~L.}\ \bibnamefont
  {Cardy}}\ and\ \bibinfo {author} {\bibfnamefont {I.}~\bibnamefont
  {Peschel}},\ }\href {\doibase 10.1016/0550-3213(88)90604-9} {\bibfield
  {journal} {\bibinfo  {journal} {Nuclear Physics B}\ }\textbf {\bibinfo
  {volume} {300}},\ \bibinfo {pages} {377} (\bibinfo {year}
  {1988})}\BibitemShut {NoStop}%
\bibitem [{\citenamefont {St{\'e}phan}\ and\ \citenamefont
  {Dubail}(2013)}]{logloverl}%
  \BibitemOpen
  \bibfield  {author} {\bibinfo {author} {\bibfnamefont {J.-M.}\ \bibnamefont
  {St{\'e}phan}}\ and\ \bibinfo {author} {\bibfnamefont {J.}~\bibnamefont
  {Dubail}},\ }\href {http://dx.doi.org/10.1088/1742-5468/2013/09/P09002}
  {\bibfield  {journal} {\bibinfo  {journal} {J. Stat. Mech P09002}\ }
  (\bibinfo {year} {2013})},\ \Eprint {http://arxiv.org/abs/1303.3633}
  {arXiv:1303.3633} \BibitemShut {NoStop}%
\bibitem [{Note2()}]{Note2}%
  \BibitemOpen
  \bibinfo {note} {Note that this is not true for the full periodic chain,
  where there are no logarithms, and the leading universal piece depends on the
  stiffness \cite {SFMP,SMP3}.}\BibitemShut {Stop}%
\bibitem [{\citenamefont {Cardy}(1991)}]{Cardy_bfield}%
  \BibitemOpen
  \bibfield  {author} {\bibinfo {author} {\bibfnamefont {J.~L.}\ \bibnamefont
  {Cardy}},\ }\href {http://dx.doi.org/10.1088/0305-4470/24/22/003} {\bibfield
  {journal} {\bibinfo  {journal} {J. Phys. A: Math. Gen.}\ }\textbf {\bibinfo
  {volume} {24}},\ \bibinfo {pages} {L1315} (\bibinfo {year}
  {1991})}\BibitemShut {NoStop}%
\bibitem [{Note3()}]{Note3}%
  \BibitemOpen
  \bibinfo {note} {I thank John Cardy for pointing that out to me.}\BibitemShut
  {Stop}%
\bibitem [{\citenamefont {Diehl}(1986)}]{Diehl}%
  \BibitemOpen
  \bibfield  {author} {\bibinfo {author} {\bibfnamefont {H.}~\bibnamefont
  {Diehl}},\ }\href@noop {} {\emph {\bibinfo {title} {{Field-theoretic approach
  to critical behaviour at surfaces}}}},\ \bibinfo {edition} {domb and
  lebowitz}\ ed.,\ Vol.\ \bibinfo {volume} {Phase transitions and critical
  phenomena}\ (\bibinfo  {publisher} {Academic Press, New-York},\ \bibinfo
  {year} {1986})\BibitemShut {NoStop}%
\bibitem [{\citenamefont {Igloi}\ \emph {et~al.}(1991)\citenamefont {Igloi},
  \citenamefont {Turban},\ and\ \citenamefont {Berche}}]{IgloiTurbanBerche}%
  \BibitemOpen
  \bibfield  {author} {\bibinfo {author} {\bibfnamefont {F.}~\bibnamefont
  {Igloi}}, \bibinfo {author} {\bibfnamefont {L.}~\bibnamefont {Turban}}, \
  and\ \bibinfo {author} {\bibfnamefont {B.}~\bibnamefont {Berche}},\ }\href
  {http://dx.doi.org/10.1088/0305-4470/24/17/012} {\bibfield  {journal}
  {\bibinfo  {journal} {J. Phys. A: Math. Gen.}\ }\textbf {\bibinfo {volume}
  {24}},\ \bibinfo {pages} {L1031} (\bibinfo {year} {1991})},\ \Eprint
  {http://arxiv.org/abs/cond-mat/0106626} {cond-mat/0106626} \BibitemShut
  {NoStop}%
\bibitem [{\citenamefont {Tsvelik}(2013)}]{Tsvelik1}%
  \BibitemOpen
  \bibfield  {author} {\bibinfo {author} {\bibfnamefont {A.~M.}\ \bibnamefont
  {Tsvelik}},\ }\href {http://dx.doi.org/10.1103/PhysRevLett.110.147202}
  {\bibfield  {journal} {\bibinfo  {journal} {Phys. Rev. Lett.}\ }\textbf
  {\bibinfo {volume} {110}},\ \bibinfo {pages} {147202} (\bibinfo {year}
  {2013})},\ \Eprint {http://arxiv.org/abs/1211.3481} {arXiv:1211.3481}
  \BibitemShut {NoStop}%
\bibitem [{\citenamefont {Iaconis}\ \emph {et~al.}(2013)\citenamefont
  {Iaconis}, \citenamefont {Inglis}, \citenamefont {Kallin},\ and\
  \citenamefont {Melko}}]{Melkoinf3}%
  \BibitemOpen
  \bibfield  {author} {\bibinfo {author} {\bibfnamefont {J.}~\bibnamefont
  {Iaconis}}, \bibinfo {author} {\bibfnamefont {S.}~\bibnamefont {Inglis}},
  \bibinfo {author} {\bibfnamefont {A.~B.}\ \bibnamefont {Kallin}}, \ and\
  \bibinfo {author} {\bibfnamefont {R.~G.}\ \bibnamefont {Melko}},\ }\href
  {\doibase 10.1103/PhysRevB.87.195134} {\bibfield  {journal} {\bibinfo
  {journal} {Phys. Rev. B}\ }\textbf {\bibinfo {volume} {87}},\ \bibinfo
  {pages} {195134} (\bibinfo {year} {2013})},\ \Eprint
  {http://arxiv.org/abs/1210.2403} {arXiv:1210.2403} \BibitemShut {NoStop}%
\bibitem [{\citenamefont {Kumano}(2013)}]{Kumano}%
  \BibitemOpen
  \bibfield  {author} {\bibinfo {author} {\bibfnamefont {Y.}~\bibnamefont
  {Kumano}},\ }\href@noop {} {\bibfield  {journal} {\bibinfo  {journal}
  {unpublished}\ } (\bibinfo {year} {2013})}\BibitemShut {NoStop}%
\bibitem [{\citenamefont {Giamarchi}(2004)}]{Giamarchi}%
  \BibitemOpen
  \bibfield  {author} {\bibinfo {author} {\bibfnamefont {T.}~\bibnamefont
  {Giamarchi}},\ }\href@noop {} {\emph {\bibinfo {title} {{Quantum Physics in
  One Dimension}}}}\ (\bibinfo  {publisher} {Oxford University Press,
  New-York},\ \bibinfo {year} {2004})\BibitemShut {NoStop}%
\bibitem [{\citenamefont {Pozsgay}(2013)}]{Pozsgay}%
  \BibitemOpen
  \bibfield  {author} {\bibinfo {author} {\bibfnamefont {B.}~\bibnamefont
  {Pozsgay}},\ }\href@noop {} {\bibfield  {journal} {\bibinfo  {journal}
  {preprint}\ } (\bibinfo {year} {2013})},\ \Eprint
  {http://arxiv.org/abs/1309.4593} {arXiv:1309.4593} \BibitemShut {NoStop}%
\bibitem [{\citenamefont {Brockmann}\ \emph {et~al.}(2014)\citenamefont
  {Brockmann}, \citenamefont {Nardis}, \citenamefont {Wouters},\ and\
  \citenamefont {Caux}}]{Caux}%
  \BibitemOpen
  \bibfield  {author} {\bibinfo {author} {\bibfnamefont {M.}~\bibnamefont
  {Brockmann}}, \bibinfo {author} {\bibfnamefont {J.~D.}\ \bibnamefont
  {Nardis}}, \bibinfo {author} {\bibfnamefont {B.}~\bibnamefont {Wouters}}, \
  and\ \bibinfo {author} {\bibfnamefont {J.-S.}\ \bibnamefont {Caux}},\ }\href
  {http://dx.doi.org/10.1088/1751-8113/47/14/145003} {\bibfield  {journal}
  {\bibinfo  {journal} {J. Phys. A: Math. Theor.}\ }\textbf {\bibinfo {volume}
  {47}},\ \bibinfo {pages} {145003} (\bibinfo {year} {2014})},\ \Eprint
  {http://arxiv.org/abs/1401.2877} {arXiv:1401.2877} \BibitemShut {NoStop}%
\bibitem [{Note4()}]{Note4}%
  \BibitemOpen
  \bibinfo {note} {The Bogoliubov transformation is $c_j^\protect \dag =\DOTSB
  \sum@ \slimits@ _k (u_{kj} d_k^\protect \dag +v_{kj}d_k)$, where the new
  fermions operators $d_k,d_k^\protect \dag $ also obey the canonical
  anticommutation rules. In case of periodic boundary conditions $k$ also
  labels momentum.}\BibitemShut {Stop}%
\bibitem [{Note5()}]{Note5}%
  \BibitemOpen
  \bibinfo {note} {In this precise case the method can be slightly improved
  using the exact result for $S_4(L,L)$ derived in the appendix \ref
  {sec:exact}, that gives a leading term $a_4=\protect \frac {2+\protect
  \qopname \relax o{ln}2}{6}$ in Eq.~(\ref {eq:infscaling}).}\BibitemShut
  {Stop}%
\bibitem [{Note6()}]{Note6}%
  \BibitemOpen
  \bibinfo {note} {Indeed, for a fixed aspect ratio $\ell /L=p/q$, we need to
  study even system sizes also multiple of $p$.}\BibitemShut {Stop}%
\bibitem [{\citenamefont {Chandran}\ \emph {et~al.}(2014)\citenamefont
  {Chandran}, \citenamefont {Khemani},\ and\ \citenamefont
  {Sondhi}}]{Sondhispectrum}%
  \BibitemOpen
  \bibfield  {author} {\bibinfo {author} {\bibfnamefont {A.}~\bibnamefont
  {Chandran}}, \bibinfo {author} {\bibfnamefont {V.}~\bibnamefont {Khemani}}, \
  and\ \bibinfo {author} {\bibfnamefont {S.~L.}\ \bibnamefont {Sondhi}},\
  }\href@noop {} {\  (\bibinfo {year} {2014})},\ \Eprint
  {http://arxiv.org/abs/1311.2946} {arXiv:1311.2946} \BibitemShut {NoStop}%
\bibitem [{Note7()}]{Note7}%
  \BibitemOpen
  \bibinfo {note} {With exotic terms such as e.g. $\left (\protect \qopname
  \relax o{ln}\protect \qopname \relax o{ln}L\right )\times \protect \qopname
  \relax o{ln}L$.}\BibitemShut {Stop}%
\bibitem [{\citenamefont {Dyson}(1962)}]{Dyson}%
  \BibitemOpen
  \bibfield  {author} {\bibinfo {author} {\bibfnamefont {F.~J.}\ \bibnamefont
  {Dyson}},\ }\href {http://dx.doi.org/10.1063/1.1703773} {\bibfield  {journal}
  {\bibinfo  {journal} {J. Math. Phys.}\ }\textbf {\bibinfo {volume} {3}},\
  \bibinfo {pages} {140} (\bibinfo {year} {1962})}\BibitemShut {NoStop}%
\bibitem [{\citenamefont {Macdonald}(1982)}]{Macdonald}%
  \BibitemOpen
  \bibfield  {author} {\bibinfo {author} {\bibfnamefont {I.~G.}\ \bibnamefont
  {Macdonald}},\ }\href {http://dx.doi.org/10.1137/0513070} {\bibfield
  {journal} {\bibinfo  {journal} {SIAM J. Math. Anal.}\ }\textbf {\bibinfo
  {volume} {13}},\ \bibinfo {pages} {998} (\bibinfo {year} {1982})}\BibitemShut
  {NoStop}%
\bibitem [{\citenamefont {Good}(1972)}]{Good}%
  \BibitemOpen
  \bibfield  {author} {\bibinfo {author} {\bibfnamefont {I.}~\bibnamefont
  {Good}},\ }\href {http://dx.doi.org/10.1063/1.1665339} {\bibfield  {journal}
  {\bibinfo  {journal} {J. Math. Phys.}\ }\textbf {\bibinfo {volume} {11}},\
  \bibinfo {pages} {1884} (\bibinfo {year} {1972})}\BibitemShut {NoStop}%
\bibitem [{\citenamefont {Affleck}(1998)}]{Affleck}%
  \BibitemOpen
  \bibfield  {author} {\bibinfo {author} {\bibfnamefont {I.}~\bibnamefont
  {Affleck}},\ }\href {http://dx.doi.org/10.1088/0305-4470/31/12/003}
  {\bibfield  {journal} {\bibinfo  {journal} {J. Phys. A: Math. Gen.}\ }\textbf
  {\bibinfo {volume} {31}},\ \bibinfo {pages} {2761} (\bibinfo {year}
  {1998})},\ \Eprint {http://arxiv.org/abs/cond-mat/9710221}
  {arXiv:cond-mat/9710221} \BibitemShut {NoStop}%
\bibitem [{\citenamefont {Bilstein}\ and\ \citenamefont
  {Wehefritz}(1999)}]{BilsteinWehefritz}%
  \BibitemOpen
  \bibfield  {author} {\bibinfo {author} {\bibfnamefont {U.}~\bibnamefont
  {Bilstein}}\ and\ \bibinfo {author} {\bibfnamefont {B.}~\bibnamefont
  {Wehefritz}},\ }\href {\doibase 10.1088/0305-4470/32/2/001} {\bibfield
  {journal} {\bibinfo  {journal} {J. Phys. A: Math. Gen.}\ }\textbf {\bibinfo
  {volume} {324}},\ \bibinfo {pages} {191} (\bibinfo {year} {1999})},\ \Eprint
  {http://arxiv.org/abs/cond-mat/9807166} {arXiv:cond-mat/9807166} \BibitemShut
  {NoStop}%
\bibitem [{\citenamefont {Wehefritz-Kaufman}(2007)}]{Wehefritz}%
  \BibitemOpen
  \bibfield  {author} {\bibinfo {author} {\bibfnamefont {B.}~\bibnamefont
  {Wehefritz-Kaufman}},\ }\href {http://dx.doi.org/10.1088/1751-8113/40/2/002}
  {\bibfield  {journal} {\bibinfo  {journal} {J. Phys. A: Math. Theor.}\
  }\textbf {\bibinfo {volume} {40}},\ \bibinfo {pages} {217} (\bibinfo {year}
  {2007})}\BibitemShut {NoStop}%
\bibitem [{\citenamefont {Gaudin}(1973)}]{Gaudin}%
  \BibitemOpen
  \bibfield  {author} {\bibinfo {author} {\bibfnamefont {M.}~\bibnamefont
  {Gaudin}},\ }\href {http://dx.doi.org/10.1051/jphys:01973003407051100}
  {\bibfield  {journal} {\bibinfo  {journal} {J. Phys}\ }\textbf {\bibinfo
  {volume} {34}},\ \bibinfo {pages} {511} (\bibinfo {year} {1973})}\BibitemShut
  {NoStop}%
\bibitem [{\citenamefont {Lazarescu}(2013)}]{LazarescuPhD}%
  \BibitemOpen
  \bibfield  {author} {\bibinfo {author} {\bibfnamefont {A.}~\bibnamefont
  {Lazarescu}},\ }\href@noop {} {Ph.D. thesis} (\bibinfo {year} {2013}),\
  \Eprint {http://arxiv.org/abs/1311.7370} {arXiv:1311.7370} \BibitemShut
  {NoStop}%
\bibitem [{\citenamefont {Nielsen}\ \emph {et~al.}(2011)\citenamefont
  {Nielsen}, \citenamefont {Cirac},\ and\ \citenamefont
  {Sierra}}]{NielsenCiracSierra}%
  \BibitemOpen
  \bibfield  {author} {\bibinfo {author} {\bibfnamefont {A.~E.~B.}\
  \bibnamefont {Nielsen}}, \bibinfo {author} {\bibfnamefont {J.~I.}\
  \bibnamefont {Cirac}}, \ and\ \bibinfo {author} {\bibfnamefont
  {G.}~\bibnamefont {Sierra}},\ }\href
  {http://dx.doi.org/10.1088/1742-5468/2011/11/P11014} {\bibfield  {journal}
  {\bibinfo  {journal} {J. Stat. Mech. P11014}\ } (\bibinfo {year} {2011})},\
  \Eprint {http://arxiv.org/abs/1109.5470} {arXiv:1109.5470} \BibitemShut
  {NoStop}%
\bibitem [{\citenamefont {Haldane}(1988)}]{Haldane}%
  \BibitemOpen
  \bibfield  {author} {\bibinfo {author} {\bibfnamefont {F.~D.~M.}\
  \bibnamefont {Haldane}},\ }\href {\doibase 10.1103/PhysRevLett.60.635}
  {\bibfield  {journal} {\bibinfo  {journal} {Phys. Rev. Lett.}\ }\textbf
  {\bibinfo {volume} {60}},\ \bibinfo {pages} {635} (\bibinfo {year}
  {1988})}\BibitemShut {NoStop}%
\bibitem [{\citenamefont {Shastry}(1988)}]{Shastry}%
  \BibitemOpen
  \bibfield  {author} {\bibinfo {author} {\bibfnamefont {B.~S.}\ \bibnamefont
  {Shastry}},\ }\href {\doibase 10.1103/PhysRevLett.60.639} {\bibfield
  {journal} {\bibinfo  {journal} {Phys. Rev. Lett.}\ }\textbf {\bibinfo
  {volume} {60}},\ \bibinfo {pages} {639} (\bibinfo {year} {1988})}\BibitemShut
  {NoStop}%
\bibitem [{\citenamefont {Bernard}\ \emph {et~al.}(1995)\citenamefont
  {Bernard}, \citenamefont {Pasquier},\ and\ \citenamefont
  {Serban}}]{BernardPasquierSerban}%
  \BibitemOpen
  \bibfield  {author} {\bibinfo {author} {\bibfnamefont {D.}~\bibnamefont
  {Bernard}}, \bibinfo {author} {\bibfnamefont {V.}~\bibnamefont {Pasquier}}, \
  and\ \bibinfo {author} {\bibfnamefont {D.}~\bibnamefont {Serban}},\ }\href
  {http://dx.doi.org/10.1209/0295-5075/30/5/009} {\bibfield  {journal}
  {\bibinfo  {journal} {Europhys. Lett.}\ }\textbf {\bibinfo {volume} {300}},\
  \bibinfo {pages} {301} (\bibinfo {year} {1995})},\ \Eprint
  {http://arxiv.org/abs/hep-th/9501044} {arXiv:hep-th/9501044} \BibitemShut
  {NoStop}%
\bibitem [{\citenamefont {Forrester}(2010)}]{Forrester}%
  \BibitemOpen
  \bibfield  {author} {\bibinfo {author} {\bibfnamefont {P.~J.}\ \bibnamefont
  {Forrester}},\ }\href@noop {} {\emph {\bibinfo {title} {{Log-gases and Random
  Matrices}}}}\ (\bibinfo  {publisher} {Princeton University Press},\ \bibinfo
  {year} {2010})\BibitemShut {NoStop}%
\end{thebibliography}%
\end{document}